%% file: main.tex
\documentclass[compsoc,conference,a4paper,10pt,times]{IEEEtran}
\IEEEoverridecommandlockouts

\usepackage{cite}
\usepackage{amsmath,amssymb,amsfonts}
\usepackage{algorithmic}
\usepackage{graphicx}
\usepackage{textcomp}
\usepackage{xcolor}
\def\BibTeX{{\rm B\kern-.05em{\sc i\kern-.025em b}\kern-.08em
    T\kern-.1667em\lower.7ex\hbox{E}\kern-.125emX}}

\usepackage{nicefrac}
\usepackage{siunitx}
\usepackage{array,framed}
\usepackage{booktabs}
\usepackage{
  color,
  float,
  epsfig,
  wrapfig,
  graphics,
  subcaption
}
\usepackage{textcomp}
\usepackage{setspace}
\usepackage{latexsym,fancyhdr}
\usepackage{enumerate}

\usepackage{xparse} 
\usepackage{xspace}
\usepackage{multirow}
\usepackage{csvsimple}
\usepackage{balance}
\usepackage{url}
\usepackage{wasysym}

\usepackage{
  tikz,
  pgfplots,
  pgfplotstable
}
\usepackage[hidelinks]{hyperref}

\usepackage{comment}

\input{code}
\usetikzlibrary{
  shapes.geometric,
  arrows,
  external,
  pgfplots.groupplots,
  matrix
}
\pgfplotsset{compat=1.9}

\usepackage{mathtools,}

\DeclareMathAlphabet{\mathcal}{OMS}{cmsy}{m}{n}

\DeclareGraphicsExtensions{%
    .png,.PNG,%
    .pdf,.PDF,%
    .jpg,.mps,.jpeg,.jbig2,.jb2,.JPG,.JPEG,.JBIG2,.JB2}

\usepackage[noindentafter]{titlesec}
\titlespacing\section{0pt}{7pt plus 4pt minus 2pt}{2pt plus 2pt minus 2pt}
\titlespacing\subsection{0pt}{7pt plus 4pt minus 2pt}{1pt plus 2pt minus 2pt}
\titlespacing\subsubsection{0pt}{7pt plus 4pt minus 2pt}{0pt plus 2pt minus 2pt}

\usepackage{soul,color}
\newcommand{\karthika}[1]{\textcolor{black}{#1}}
\newcommand{\roberto}[1]{\textcolor{red}{[(Roberto): #1]}}

\begin{document}


\title{Categorizing Service Worker Attacks and Mitigations}

\author{\IEEEauthorblockN{Karthika Subramani}
\IEEEauthorblockA{\textit{University of Georgia}\\
ks54471@uga.edu}
\and
\IEEEauthorblockN{Jordan Jueckstock}
\IEEEauthorblockA{\textit{North Carolina State University}\\
jjuecks@ncsu.edu}
\and
\IEEEauthorblockN{Alexandros Kapravelos}
\IEEEauthorblockA{\textit{North Carolina State University}\\
akaprav@ncsu.edu}
\and
\IEEEauthorblockN{Roberto Perdisci}
\IEEEauthorblockA{\textit{University of Georgia}\\
perdisci@uga.edu}
}

\maketitle

\input{abstract}



\input{intro}    
\input{background}

\input{sok_abuse}
\input{sw_mitigations}
\input{open_problems_mitigations}
\input{evaluation}
\input{policies_implementation}

\input{relatedwork}
\input{conclusion}


\bibliographystyle{./bibliography/IEEEtran}
\bibliography{references}
%

\appendix
\input{appendix}

\end{document}

%% file: code.tex
\usepackage{listings} 
\usepackage{color} 
\definecolor{mygreen}{rgb}{0,0.6,0}
\definecolor{mygray}{rgb}{0.5,0.5,0.5}
\definecolor{mymauve}{rgb}{0.58,0,0.82}
 
\definecolor{darkgray}{rgb}{.4,.4,.4}
\definecolor{purple}{rgb}{0.65, 0.12, 0.82}
 
\lstdefinelanguage{JavaScript}{
keywords={typeof, new, true, false, catch, function, return, null, catch, switch, var, if, in, while, do, else, case, break},
keywordstyle=\color{blue}\bfseries,
ndkeywords={class, export, boolean, throw, implements, import, this},
ndkeywordstyle=\color{darkgray}\bfseries,
identifierstyle=\color{black},
sensitive=false,
comment=[l]{//},
morecomment=[s]{/*}{*/},
commentstyle=\color{olive}\ttfamily,
stringstyle=\color{red}\ttfamily,
morestring=[b]',
morestring=[b]"
}

\lstdefinelanguage{C}{
keywords={FROM_HERE, Start, SetIdleDelay, },
keywordstyle=\color{blue}\bfseries,
ndkeywords={class, export, boolean, throw, implements, import, this},
ndkeywordstyle=\color{darkgray}\bfseries,
identifierstyle=\color{black},
sensitive=false,
comment=[l]{//},
morecomment=[s]{/*}{*/},
commentstyle=\color{purple}\ttfamily,
stringstyle=\color{red}\ttfamily,
morestring=[b]',
morestring=[b]"
}
 

%% file: abstract.tex
\begin{abstract}
Service Workers (SWs) are a powerful feature at the core of {\em Progressive Web
Apps}, namely web applications that can continue to function when the user's
device is offline and that have access to device sensors and capabilities
previously accessible only by native applications. 
During the past few years, researchers have found a number of ways in which SWs
may be abused to achieve different malicious purposes. For instance, SWs may
be abused to build a web-based botnet, launch DDoS attacks, or perform
cryptomining; they may be hijacked to create persistent cross-site scripting
(XSS) attacks; they may be leveraged in the context of side-channel attacks to
compromise users' privacy; or they may be abused for phishing or social
engineering attacks using web push notifications-based malvertising.

In this paper, we reproduce and analyze known attack vectors related to SWs and
explore new abuse paths that have not previously been considered. We systematize
the attacks into different categories, and then analyze whether, how, and
estimate when these attacks have been published and mitigated by
different browser vendors. Then, we discuss a number of open SW security
problems that are currently unmitigated, and propose SW behavior monitoring
approaches and new browser policies that we believe should be implemented by
browsers to further improve SW security. Furthermore, we implement a
proof-of-concept version of several policies in the Chromium code base, and also
measure the behavior of SWs used by highly popular web applications with respect
to these new policies. Our measurements show that it should be feasible to
implement and enforce stricter SW security policies without a significant impact
on most legitimate production SWs.
\end{abstract}

%% file: intro.tex
\section{Introduction}
\label{sec:intro}

Service Workers~\cite{sw_intro} are a {\em powerful
feature}~\cite{sw_power}
at the core of {\em Progressive Web
Apps}~\cite{pwa_intro}, namely web
applications that can continue to function when the user's device is offline and
that have access to device sensors and capabilities previously accessible only
by native applications. In practice, a Service Worker (SW) is a JavaScript
Worker~\cite{sw_intro} script with the following high-level properties: (i) it is installed by a web application rendered in a browser; (ii) after installation,
the SW can act as a proxy for network requests issued by its web application,
and can thus control how web content is retrieved (e.g., from a local cache or
the network) and what content is eventually passed to the application; (iii) it
is an event-driven process that runs in the background, even when its web
application is not actively rendered on the browser, and that can be activated
by the browser based on events such as receiving a web push
message~\cite{push_api}
or a request to fetch a web page on behalf of its web application, among others.

Because SWs are a powerful feature, browser developers are mindful of potential
security risks that come with them. Therefore, over time browsers have
implemented a number of security policies around SWs, to limit potential abuse
(see Section~\ref{sec:background}). As an example, SW files can only be
requested from a secure first-party origin (essentially, via HTTPS and from the
same domain as the installing web application's origin). However, during the
past few years, researchers have found a number of ways in which SWs may still
be abused to achieve different malicious purposes. For instance, SWs may be
abused to build a web-based botnet~\cite{masterofweb}, launch DDoS attacks, or
perform cryptomining~\cite{pridepwa}; they may be {\em hijacked} to create
persistent cross-site scripting (XSS) attacks~\cite{Chinprutthiwong2020}; they
may be leveraged in the context of side-channel attacks to compromise users'
privacy~\cite{karami2021awakening}; or they may be abused for
phishing~\cite{pridepwa} or social engineering attacks using web push
notifications-based malvertising ~\cite{subramani2020}.

In this paper, we reproduce and analyze known attack vectors related to SWs, and
explore new abuse paths that have not previously been considered
(Section~\ref{sec:attacks}). We first systematize this information by grouping
the attacks into different categories, based on the fundamental SW security
weaknesses that make the attacks possible. Afterwards, we analyze whether, how,
and estimate when these attacks have been published and mitigated by different
browser vendors, and organize this information into an {\em attacks and
mitigations} timeline (see Section~\ref{sec:mitigations},
Table~\ref{tab:sok_attacks} and Figure~\ref{fig:attack_timeline}). Then, we
discuss a number of open SW security problems that to the best of our knowledge
are currently unmitigated. Accordingly, we propose SW behavior monitoring
approaches and new browser policies that we believe should be implemented by
browsers to further improve SW security (Section~\ref{sec:open_problems}). While
preventing all types of SW abuse may not be possible, we aim to propose policies
that can limit the damage that potential SW attacks can make, while minimizing
the impact the proposed browser changes may have on existing legitimate SW code.
To demonstrate the feasibility of the proposed browser policy changes, we
implement a proof-of-concept version of several policies in the Chromium code
base, and also measure the behavior of SWs used by highly popular web
applications with respect to these new policies (Sections~\ref{sec:measurements}
and~\ref{sec:policy_impl}).

In summary, we make the following main contributions:
\begin{itemize}
    \item We reproduce previously known attacks that abuse SWs, discuss new
    paths for abuse that were previously not considered, and systematize these
    SW attacks into categories based on the fundamental features that make them
    possible. 
    \item We study whether, how, and estimate when the SW attacks have been
    mitigated by different browser vendors, and organize this information in an
    {\em attacks and mitigations} timeline.
    \item We discuss open security problems related to SWs and propose new
    browser policies that aim to reduce the potential for future SW abuse.
    \item Finally, we implement a proof-of-concept version of a number of such
    policies in Chromium. Also, we measure how policy parameters could be tuned
    to limit SW abuse without significantly impacting legitimate SWs used by
    popular websites. To this end, we build a {\em SW forenic engine}, namely an
    instrumented Chromium browser that allows us to obtain fine-grained
    information on the behavior of SW code for real-world web applications.
    \item In addition, we disclosed the new attacks we found to browser vendors
    and obtained confirmation of their effectiveness. We will release our
    repository of reproduced and new SW attacks, SW forensic engine,
    proof-of-concept browser policy implementations, and measurement results to
    the community (the entire repository will be released after publication; the
    code and demos of our collection of attacks are available
    at~\cite{attacks_repo}).
\end{itemize}

%% file: background.tex
\section{Background}
\label{sec:background}
In this section, we provide a brief background on Service Workers (SWs), focusing primarily on properties that are used as part of the attacks and mitigations described in later sections of this paper. 

\subsection{Service Workers}

A Service Worker (SW) is a JavaScript Worker~\cite{sw_intro}, namely an
event-driven script that runs in the background and that does not have direct
access to the DOM. To run in the background, a SW first needs to be registered
by a web page. The SW code has to be contained in a JavaScript file hosted under
the same origin as the origin of the web page that invokes its registration.
Once installed, the SW can be programmed to cache web pages that may be later
served to the user even if the browser is offline. This allows a web application
running in the browser to behave more like a native application, even when the
user's device connectivity to the Internet is unreliable. In addition, SWs can
receive push messages and send web push notifications to the user even when the
related web application is not open on the browser, in a way similar to native
application's notifications.

\subsection{SW Lifecycle}
Once a website registers a SW, the SW code goes through an installation and
activation phase, after which it can {\em control} web page requests under the
website's origin.Before installation completes, the SW can import additional
scripts into the worker's context by using the importScripts API. As such,
additional code may be imported from any third-party origin. The SW is ready to
use only after it is activated. Installed SWs can be updated at any point of
time to a new version. Automatic checks for these updates are scheduled by the
browser at an interval of 24 hours or whenever a user visits a web page that the
SW controls. An update could also be triggered at any point of time by using the
{\em Update} API. Furthermore, a SW can be explicitly de-registered by its web
application. 

Once installed, the SW is activated immediately, if there is no pre-existing SW
installed from the same origin. Otherwise, it needs to wait for a previously
installed SW to finish its execution. If required, this wait period can be
skipped by using the {\em skipWaiting} API. Once activated, the status of the SW
remains set to {\em running} until it is terminated by the browser. Each time an
event is sent to the SW, the browser activates the SW code and signals the SW
about the event.

%

\subsection{SW Scope}
Each SW has a {\em scope} that can be specified during the registration process~\cite{sw_reg}. The scope represents the URL path under which web pages are controlled by the SW\footnote{For instance, a SW registered under origin \url{https://example.com} with scope \url{/test} has control over all web content requests under \url{https://example.com/test}.}. If no scope is specified, then by default the SW acquires the scope of the URL path under which the SW file is hosted.
%
%
Currently, a website can have only one SW registered with a given scope. However, multiple SWs can be registered under the same origin if they have different scopes. If a SW, $SW_R$, is registered with a scope at the root level (i.e., {\tt scope=`/'}), it will gain control over all pages of the website. However, if a second SW, $SW_A$, is registered with a more specific scope (e.g., {\tt scope=`/test'}), this SW is given priority over pages under its specific scope. Therefore, any requests made for web pages under this specific scope (e.g., \url{/test/page.html}) will be handled by $SW_A$ and not $SW_R$. However, $SW_A$ will not have access to requests made by web pages outside its scope. 

Notably, a user doesn't have to visit a web page within the scope of the SW for that particular SW to be registered. For example, when the user visits a web page at the website's root level (e.g., \url{/index.html}), that page can register multiple SWs with different scopes. 

\subsection{Handling Network Requests}
Once a SW is activated, it can listen to {\em fetch} events from web pages under
its scope and thus intercept requests for web content. The SW can then make
network requests for the requested content and cache them (using the {\em Cache}
API). Later, when a cached resource is requested again, it can be served from
the cache, which can help to reduce content load latency and enables a web
application to continue working even if the device is offline. As a result, SWs
gain a powerful ability that allows them to monitor users' requests and also
modify the response sent back to the web page. As we will discuss in later
sections, this ability could lead to SW abuse and potential leaks of sensitive
information to third party sites (see Sections~\ref{sec:attacks}
and~\ref{sec:open_problems}).

\subsection{Push Notifications}
A significant component of SWs is the ability to send web push notifications to users who grant permission. To use push notifications, a SW has to subscribe to a push service by using the {\em PushManager.subscribe} API\cite{push_manager_api}. This includes adding an {\em applicationServeKey} to the options. Once subscription is successful, the browser creates an endpoint URL and an {\em auth} secret key\cite{push_subscription_api} that shouldn't be shared outside the application. These details are later used to steer push messages to the correct SW. Whenever, a push message is sent to the browser, the browser activates the corresponding SW and signals a {\em push} event that the SW can handle. More details about subscribing to push notifications can be found in~\cite{push_subscription_api}.

While push messages are received in the background, SWs can also ask the browser to display a visual notification (typically in response to a push message) to the user. To this end, SWs can call {\em showNotification} to display a message on the user interface. Notice that while push messages and notifications are typically used together, SWs may choose not to call {\em showNotification} in response to a push message being received (in some browsers, this will trigger a default notification message issued by the browser itself). Similarly, a SW can call {\em showNotification} independently from receiving a push message.

To send notifications to the user, a SW needs to request a one-time explicit user consent (usually during the SW registration phase), by invoking {\em Notification.requestPermission()}. However, in addition to the user granting permission to the SW via the browser UI, the browser itself can only display visual notifications to the user if OS-level permission is granted. Different OSes have their own policies regarding how applications (including the browser) can obtain such permission. As an example, in case of MacOS the permission is disabled by default, unless the user specifically grants the permission (for instance, at the end of the browser software installation process). On the contrary, Windows grant such permission by default.

\subsection{Periodic Background Sync} 
The {\em Periodic Background
Sync}\cite{periodic_sync_api}
API  allows web applications to configure their SWs so to make updates in the
background at a periodic time interval. It can be used to trigger {\tt
periodicsync} events from the SW without any event being received from a remote
server. 
Effectively, this feature allows a web application to keep its SW and cached
content up to date. This API is currently supported by Chrome and other
Chromium-based browsers, such as Edge and Opera. However, given its ability to
operate in the background, the {\em Periodic Background Sync} poses a potential
security threat that has refrained other browsers, such as Firefox and Safari,
from implementing it~\cite{sync_concerns}. To curb its possible abuse, browsers
need to enforce a number of restrictions on the API use
\cite{periodic_sync_rules, periodic_sync_security,sync_issue}.

\subsection{Security Policies}
In general, browsers enforce a number of default security policies to limit
potential SW abuse. For instance~\cite{sw_policies,sw_security}:
\begin{enumerate}
    \item Only secure origins (HTTPS sites) can register SWs.
    \item The JavaScript file containing SW code must be hosted under the same
    origin as the website that registers the SW.
    \item A SW should be terminated if the SW code has been idle for more than 30
    seconds or if an event takes more than 5 minutes to process.
    \item Push notifications should trigger a user-visible notification if the
    SW does not explicitly issue one.
    \item The use of some APIs (e.g., {\em Periodic Background Sync}) should be restricted by permissions that must be granted by the browser (not necessarily via a direct UI request to the user\cite{periodic_sync_rules}).
\end{enumerate}
Unfortunately, not all browsers implement all policies and, when implemented,
differences exist among browser vendors. In the rest of the paper, we discuss both previously known and new
ways in which an attacker could still abuse SWs to achieve malicious goals despite the SW constraints listed above.




%% file: sok_abuse.tex
\section{Service Worker Abuse}
\label{sec:attacks}

\input{tables/sok_attacks}

In this section, we describe and categorize a number of attacks that can be
launched by abusing Service Workers (SWs) in different ways. We group the
attacks into categories based on the root SW features that make them possible.
For most of the attacks we discuss, we (re-)produce our own proof-of-concept
implementations, which we tested on a large number of browser versions from five major browser vendors and have shared them publicly~\cite{attacks_repo} (except for two new attacks that we disclosed to vendors but are not yet mitigated; see Section~\ref{sec:ethical}).
A summary of the attacks we consider is provided in Table \ref{tab:sok_attacks},
which includes a reference to relevant previous publications or online resources
in which an attack was described. To the best of our knowledge, some of the
attack variants we discuss were not previously considered and are thus marked as
``New.'' 
In addition, Table~\ref{tab:sok_attacks} provides detailed information on
different browser features or APIs that are exploited for each attack and
information about what browser versions were first affected and what version
provided a fix, if any. In this section we focus on categorizing the attacks,
whereas browser mitigations are discussed in Section~\ref{sec:mitigations}.

\subsection{Continuous Execution}
\label{sec:cont_exe}
Because SWs can run in the background, can issue network requests, and {\em can
be activated at any time} (even when their related website origin is closed),
they can be abused to {\em stealthily} run unwanted or malicious code. For
instance, SWs could be abused by a malicious website to run {\em cryptomining}
code~\cite{pridepwa} or to build a {\em web-based botnet}~\cite{masterofweb}.
Such types of attacks are generally enabled by artificially prolonging the
amount of execution time granted by the browser to SW code running in the
background, thus (approximately) achieving {\em continuous execution}.  

\vspace{3pt}
\noindent \textbf{WebBot:} Papadopoulos et al.~\cite{masterofweb} describe how
to build a SW-based botnet. If a victim visits a malicious website $M$, this
website can register a SW, $S_M$, which can run in the background. When
executed, $S_M$ could implement code that (i) reaches out to a
command-and-control (C\&C) website to receive commands and (ii) execute the
received command to perform actions such as participating in DDoS attacks, {\em
distributed password cracking}, function as a {\em relay proxy}, etc. 

For the botnet to function properly, $S_M$ needs to be periodically
(frequently) activated. Papadopoulos et al.~\cite{masterofweb} mention that this
would be possible by using the {\em BackgroundSync} API \cite{sync_manager_api}.  Based on this information alone, we were
initially unable to fully reproduce the attack. However, we found an online
discussion about the attack by Chromium developers~\cite{sync_issue}, which
stated that the attack was possible primarily due to a bug that allowed the {\em
Update} API (see Section~\ref{sec:background}) to be invoked from a SW's {\em
activate} listener. By combining the information provided in \cite{masterofweb}
and \cite{sync_issue}, we were able to reproduce the attack as explained below.
For the attack to work, the following components are required:
    \begin{itemize}
        \item SW script whose content keeps changing at server side to appear fresher than the SW already registered. 
        \item Leveraging the {\em Update} API, which checks if there are any changes made to the SW file and fetches the updated version of the SW script from the server.
        \item Leveraging the {\em BackgroundSync} API to activate the SW every time the browser is re-opened.
    \end{itemize}

\begin{lstlisting}[caption={Example of SW self-update on {\em activate}},language=JavaScript, label=code:sync_update]
function wait(ms) {
  const tmp = setInterval(() => { /* do bad */ }, 100);
  return new Promise(res => setTimeout(res, ms)); 
}
self.addEventListener('activate', event => {
  self.registration.sync.register('foo');
  // Wait < 30s
  event.waitUntil(wait(25000).then(() => {
    self.registration.update();  }));
});
\end{lstlisting}

As shown in Listing~\ref{code:sync_update}, as the SW is activated it registers
a {\em BackgroundSync} and then calls the {\em Update} API after a predefined
timeout. In general, the SW execution is supposed to terminate after a fixed
period of time (a few minutes). Since the SW file in the server keeps changing,
calling the {\em update} method will fetch the newer version of the SW script.
This action is followed by the browser raising the {\em activate} event, which
causes the corresponding listener function in the SW code to be executed, where
malicious code can be invoked. This cycle repeats, until the browser is closed
(or the SW is explicitly unregistered). When the browser is re-opened, the {\em
BackgroundSync} triggers a sync event and the SW will be activated again
restarting the execution cycle.  

%

\vspace{3pt}
\noindent \textbf{PushExe:} Lee et al.~\cite{pridepwa} demonstrated that if an attacker was successful in registering a SW and obtaining push notification permission from the user, she could then leverage the {\em Push} API to activate the SW code at any moment. In some browsers, the attack could be rendered stealthy if the SW code does not explicitly invoke the {\em showNotification} API when a push notification is received. Using this approach, the authors were able to keep the SW running continuously in the background for long periods of time, for instance to perform {\em cryptomining}.

This attack was found to work in Firefox, Edge, and the UC Browser~\cite{pridepwa}, though the attack is not stealthy in Chrome because the browser displays a default notification message for every push event, which may alert the user about the presence of a malicious SW running in the background.

By independently reproducing and testing this attack, we verified that in Firexfox and Edge the browser revokes the push subscription of a SW (i.e., the SW cannot receive new push events), if the SW fails to show a notification after receiving a push message for 15 and 3 times respectively, thus blocking the attack, as also mentioned in~\cite{pridepwa}. 
However, the attack could still be made continuously stealthy in Firefox (whereas Edge does not appear to be affected) by simply renewing the SW registration in the background, after a few push messages are received (i.e., before exceeding the browser's limit for ``silent'' push events), as shown in Listing~\ref{code:show_not_code} (in Appendix).

\vspace{3pt}
\noindent \textbf{[New] StealthierPushExe:} While working to attain SW Continuous Execution, we discovered a variant of {\bf PushExe} that can overcome the limitation of default notifications being displayed to the user, which would then prevent potentially alarming the user of suspicious activity.
Further, this would allow for frequently activating the SW in a stealthier way (i.e., with no visible UI signal), making it possible to achieve stealthy continuous execution. As explained in section~\ref{sec:ethical}, after our disclosure,  Chromium developers are in the process of patching this issue and therefore, we haven't included the details of the attack in this version of the paper. Once it is resolved, we will update the paper with more details.  

We verified that this new attack work on both desktop (Windows 10) and
mobile (Android 11) devices, and have disclosed it to the Chrome developers (see
Section~\ref{sec:ethical}).


\subsection{Side-Channels}
\label{sec:side-channels}

This category of abuse includes attacks that allow unauthorized parties to
leverage SWs to gain sensitive information by bypassing browser isolation. 
%

\vspace{3pt}
\noindent \textbf{OfflineOnload:} In~\cite{pridepwa}, Lee et al. propose a history-sniffing attack that works as follows. A user first visits the attacker's website, which registers a SW. At a later time, if the user again opens  the attacker's website in {\em offline mode}, the SW will intercept the request and return a page that includes a number of {\tt iframe}s whose URL points to third-party target sites. The attacker's goal is to determine if the user previously visited those sites.   
Lee et al. found that in some browser versions, such as Firefox 59.0.2 and Safari 11.1, if the browser is in offline mode, the top page (i.e., the attacker's page) is sent an {\em onload} event related to an embedded {\tt iframe} only if the target site had already been visited by the user and a corresponding SW (with offline support) had been registered. Therefore, the attackers can register an {\em onload} event handler to sense if a third-party site embedded in an {\tt iframe} was previously visited by the user.

\vspace{3pt}
\noindent \textbf{PerformanceTiming:} In a recent paper by Karami et
al.~\cite{karami2021awakening}, the authors propose two different
history-sniffing attacks. Both approaches involve the user visiting the
attacker's website, which includes an {\em iframe} that loads content from a
third-party target site. Also, the attacks assume that the target website was
previously visited by the user, and that it registered a SW. Furthermore, the
{\em iframe}'s source URL must fall within the scope of the targeted website's SW. 

The first attack ({\bf PerformanceTiming1}) identifies the presence of a
previously registered third-party SW by monitoring two attributes of {\em
PerformanceResourceTiming} API, namely {\em workerStart} and {\em
nextHopProtocol}. The values of these attributes change depending on whether the
resource is being loaded when the target page request is served via a SW,
compared to when no SW is yet registered, and can thus be used to infer whether
the page was previously visited by the user. While working to reproduce this
attack, we additionally found that in Firefox there exists another property of {\em PerformanceResourceTiming}, called {\em initiator}, that can also be used
to identify the presence of a SW in a similar way.

The second attack ({\bf PerformanceTiming2}) is a timing-based side-channel
attack that measures the loading time for the requested {\tt iframe} resource on
the user's machine, which can be compared to a pre-calculated loading time of
the resource without the presence of a SW. Because SWs often cache resources to
optimize performance and enable offline browsing, the difference in the loading
times can help determine the presence of a SW~\cite{karami2021awakening}. 

\subsection{SW Hijacking}
\label{sec:hijacking}

We now discuss attacks that involve {\em hijacking} SW functionalities,
by either injecting malicious code into a legitimate SW or by injecting a
malicious SW into a benign origin. 

\vspace{3pt}
\noindent \textbf{XSS:} In~\cite{Chinprutthiwong2020}, Chinprutthiwong et al.
present an XSS attack that can be used to hijack a legitimate
site's SW. They found that the URL path of a SW script can in some cases be
manipulated to inject an attacker's script into the SW code. This is possible
because some websites use dynamic URL query parameters in the SW's URL path that
depend on the {\em window.location} API. The authors demonstrate that the
attacker could modify the URL parameters by tricking the users to visit a
carefully crafted target URL. Although the user ends up visiting the legitimate
target website, failure to validate the URL parameters could result in the
injection of attacker-controlled code into the SW context during the
registration of a legitimate SW. Such an attack is stealthy in that it would go
r by the user or the targeted website. 

\vspace{3pt}
\noindent \textbf{[New] ExtensionHijack:} We discovered another possible
approach to hijack a legitimate website's SW. Specifically, we found that
browser extensions can be used to inject malicious SW code in the scope of any
benign origin.

Specifically, Firefox is vulnerable to SW hijacking by extensions because they are
allowed to use the {\em FilterResponse} API, which enables them to modify the
request made to fetch a SW script file during its registration phase. This API
is unique to Firefox and we leverage it to demonstrate this new attack.

To this end, we developed a basic Firefox extension that has the capability to
intercept requests using {\em WebRequest} and {\em WebRequestBlocking}
permissions, which are commonly used by popular extensions, including ad
blockers. Next, we need to filter requests  made for obtaining the SW script.
To achieve this, our extension uses the {\em OnBeforeSendHeaders} API to
intercept requests and obtain their HTTP headers. We identify SW script
requests, as well as scripts that are imported by the SW, by looking for the
header parameter {\em name} with value {\em Service-Worker}. Once a request for
SW code is identified, we read the SW file's content using the API {\em
FilterResponse}. Before sending the file's data, we can insert at the beginning of the file a malicious code snippet (although in our
proof-of-concept extension we inject harmless code) as shown in
Listing~\ref{code:ext_code} (in Appendix). 

The advantage of this attack is that the extension itself does not explicitly
execute malicious code. Rather, the extension uses allowed APIs to inject
additional code to be executed in the context of a SW. This may make it more
difficult for extension stores to classify the extension itself as malicious in
the first place. Additionally, because malicious extensions often go unnoticed
for long periods of time~\cite{pantelaios2020you}, the impact of this attack may
be significant. Even if the extension is detected as malicious and removed from
the browser after installation, it may be too late, as the extension may have
already injected malicious SW code under many highly popular website origins,
which will continue to execute on a potentially large number of browsers even
after the extension is removed from the store and the browser, until the SW code is updated. 

\vspace{3pt}
\noindent \textbf{[New] LibraryHijack:} 
Website owners can leverage third-party libraries from ``push providers'' (e.g.,
OneSignal.com, SendPulse.com, iZooto.com, etc.) to conveniently enable and
manage push notification campaigns. Typically, this entails including
third-party code to run within the SW of a website, $W$. As a result, the
provider of the third-party code gains complete access to $W$'s SW, whose
capabilities go much beyond providing push notifications. For example, the SW
script could be modified to intercept all fetch requests and inject new page
content that may harvest sensitive user information and relay it back to an
unauthorized server. Currently, there are no restrictions posed on
functionalities of third-party SW libraries, and in
Section~\ref{sec:measurements} we discuss our findings on a third-party library
that indeed seems to misuse imported push service code to track all web pages
visited by the user on $W$.

\subsection{Other Attacks}
\label{sec:social_eng}

SWs can also be abused to launch social engineering attacks by presenting  misleading or malicious web push notification (WPN) messages to users. In both the attacks described below, a website must first register a SW and then obtain permission from the user to send notifications.
Unfortunately, social engineering attacks are difficult to mitigate directly through browser policies and restrictions on SWs. Rather, such attacks typically require content analysis, such as an analysis of the information displayed by WPN, what sites they lead to when clicked on, etc.~\cite{subramani2020}. Therefore, we discuss these attacks only briefly below and do not include them in Table~\ref{tab:sok_attacks}.
Other web attacks that are partly related to using SWs are also discussed in
Section~\ref{sec:relatedwork}.

\vspace{3pt}
\noindent \textbf{Phishing:} Lee et al.~\cite{pridepwa} discussed the
possibility of launching phishing attacks via WPNs. For instance, a malicious SW
could issue a notification that displays the Chrome icon and a message such as
``Google Chrome Premium,'' and a ``DOWNLOAD'' button, which when clicked on
could lead the user to installing malicious code. Furthermore, the authors
discuss how in some cases an attacker could extract the {\em PushSubscription}
object from network traffic~\cite{pridepwa}, and then use it to spoof push
messages as arriving from a legitimate domain. 
%

\vspace{3pt}
\noindent \textbf{Malvertising:} Although web push notifications (WPNs) were
initially meant to be used to send first-party messages to users to keep them
engaged with a website's own content, WPNs have since become an an alluring
platform for advertisers to reach users even when a given publisher website is
not being visited. For instance, ad networks such as VWO Engage (formerly
PushCrew), Roost, PushAd, etc., provide software that allows web developers to
easily include WPN-based ads to their websites. To this end, web developers
typically include third-party SW code provided by these companies to their
websites. Besides potentially exposing a website's SW to the LibraryHijack
attack described earlier, the website may also be responsible for exposing
users to malicious ads via their WPNs, as reported in~\cite{subramani2020}.

\subsection{Ethical Considerations and Disclosure}
\label{sec:ethical}
All attacks were tested using our own test websites and lab client machines. No
real user or production website was affected by our tests. 

We disclosed our findings to affected browser vendors. First, on
April 23rd, 2021, we reported the StealthierPushExe attack (see
Section~\ref{sec:cont_exe}) to the Chrome, Opera and Edge developers. After
about one month, we received confirmation that the attack affects Chromium-based
browsers (i.e., including Chrome, Opera, Edge, and likely several other less
popular browsers). The Chromium developers are currently discussing (in a
private online forum) possible fixes. Some of the steps being discussed for
patching the issue follow an approach similar to some of the recommendations we
propose in this paper for restricting SW execution (notice that we developed our
proposed mitigation ahead of disclosing the attack to the Chromium team). To the
best of our knowledge (at the time or writing), the StealthierPushExe attack has
not yet been mitigated. Therefore, our public release of the collection of SW
attacks~\cite{attacks_repo} discussed in this paper excludes the StealthierPushExe attack.

Furthermore, we also disclosed the Extension Hijack attack (see
Section~\ref{sec:hijacking}) to the Firefox developers in June 2nd,
2021, who confirmed that Firefox is indeed still vulnerable to this attack. The
developers are still discussing removing extensions' access to network requests
related to SW code and its import scripts. As for the Library Hijack attack, we
have not disclosed it to browser vendors because this attack relates to how
service workers are often misunderstood and misused by web developers, rather
than being a browser vulnerability per se. However, we plan to inform browser
vendors of this attack and the mitigation proposed in our paper once it is
accepted for publication.

%% file: tables/sok_attacks.tex

\begin{table*}[htpb]
\setlength{\belowcaptionskip}{-20pt} 
\setlength{\footskip}{30pt}
\setlength{\abovecaptionskip}{5pt plus 3pt minus 2pt} 
\caption{Overview of attacks and impacted browser versions. Legend: (\CIRCLE) first attack impact; (\Circle) fix released; (\LEFTcircle) partial fix released; (\XBox) no fix released yet; (\davidsstar) attack not possible.} 
\centering
\label{tab:sok_attacks}
\scriptsize
\resizebox{\linewidth}{!}{%
\begin{tabular}{c|c|c|c|c|c|c|c|c|c|c|c|c|c|c|}
\toprule
\hline
 &  & \multicolumn{8}{c|}{{\bf Abuse Vectors}} & \multicolumn{5}{c|}{{\bf Browsers}} \\ \cline{3-15} 
\multirow{-5}{*}{\begin{tabular}[c]{@{}c@{}} {\bf SW Abuse} \\ {\bf Categories} \end{tabular}} & \multirow{-5}{*}{\begin{tabular}[c]{@{}c@{}} {\bf Attacks} \end{tabular}}
& \rotatebox{90}{Push API} & \rotatebox{90}{Notification API} & \rotatebox{90}{Sync API} & \rotatebox{90}{Performance API~~} & \rotatebox{90}{Update API}  & \rotatebox{90}{ImportScripts} & \rotatebox{90}{\begin{tabular}[c]{@{}l@{}}{\em iframe} inclusion\end{tabular}} & \rotatebox{90}{\begin{tabular}[c]{@{}l@{}}3rd-party code\end{tabular}} 
& \multirow{-4}{*}{\begin{tabular}[c]{@{}c@{}}{Chrome}\end{tabular}} &  \multirow{-4}{*}{\begin{tabular}[c]{@{}c@{}}{Firefox}\end{tabular}} &  \multirow{-4}{*}{\begin{tabular}[c]{@{}c@{}}{Edge}\end{tabular}} &  \multirow{-4}{*}{\begin{tabular}[c]{@{}c@{}}{Safari}\end{tabular}} &  \multirow{-4}{*}{\begin{tabular}[c]{@{}c@{}}{Opera}\end{tabular}} \\ \hline

    & \begin{tabular}[c]{@{}c@{}}WebBot\\~\cite{masterofweb} \end{tabular} 
    &  &  & \checkmark  &   & \checkmark  &  &  & 
    & \begin{tabular}[c]{@{}c@{}}\CIRCLE \space  v69.0\\ \Circle \space  v70.0 \end{tabular} & \begin{tabular}[c]{@{}c@{}} \CIRCLE \space v57.0\\ \LEFTcircle \space v60.0\end{tabular} &  \begin{tabular}[c]{@{}c@{}} -\\ \Circle \space v80.0\end{tabular}&  \davidsstar  & \begin{tabular}[c]{@{}c@{}} \CIRCLE \space v56.0\\ \Circle \space v57.0\end{tabular} \\ \cline{2-15}
   
    & \begin{tabular}[c]{@{}c@{}}PushExe\\~\cite{pridepwa} \end{tabular} 
    & \checkmark  &  &  &  &  &  &  & 
    & \davidsstar & \begin{tabular}[c]{@{}c@{}}\CIRCLE \space v59.0\\ \XBox \end{tabular} 
    &  \davidsstar & \davidsstar & \davidsstar  \\ \cline{2-15} 
    
\multirow{-6}{*}{{\begin{tabular}[c]{@{}c@{}} Continuous \\ Execution\end{tabular}}} 

    & \begin{tabular}[c]{@{}c@{}}StealthierPushExe\\ {[New]}\end{tabular} & \checkmark  & \checkmark &  &  &    &  &  &  
    & \begin{tabular}[c]{@{}c@{}} \CIRCLE \space v85.0 \\ \XBox \end{tabular}
    & \begin{tabular}[c]{@{}c@{}}\davidsstar \end{tabular}
    & \begin{tabular}[c]{@{}c@{}}\CIRCLE \space v.85.0 \\ \XBox  \end{tabular}
    & \davidsstar & \begin{tabular}[c]{@{}c@{}}\CIRCLE \space v71.0\\ \XBox \end{tabular} \\ \hline
    
    & \begin{tabular}[c]{@{}c@{}}OfflineOnload\\\cite{pridepwa} \end{tabular} 
    &  &  &  &  &  &   & \checkmark  &    
    & \davidsstar & \begin{tabular}[c]{@{}c@{}}\CIRCLE \space v59.0 \\ \XBox \end{tabular} & \davidsstar 
    & \begin{tabular}[c]{@{}c@{}}\CIRCLE \space v11.1 \\ {\XBox} \end{tabular} & \davidsstar \\ \cline{2-15} 

\multirow{-2}{*}{\begin{tabular}[c]{@{}c@{}}Side-Channels\end{tabular}} & 

    \begin{tabular}[c]{@{}c@{}}PerformanceTiming1\\~\cite{karami2021awakening}\end{tabular} 
    &  &  &  & \checkmark   &  & & \checkmark  &
    & \begin{tabular}[c]{@{}c@{}}\CIRCLE \space v79.0 \\ \Circle \space  v83.0 \end{tabular} & \begin{tabular}[c]{@{}c@{}}\CIRCLE \space v72.0 \\ \LEFTcircle \space  v73.0 \end{tabular} 
    &  \begin{tabular}[c]{@{}c@{}} \CIRCLE \space v79.0 \\\Circle \space  v83.0 \end{tabular}  & \begin{tabular}[c]{@{}c@{}}\CIRCLE \space v12.1 \\ \LEFTcircle \space v14.0 \end{tabular}  
    &  \begin{tabular}[c]{@{}c@{}} \CIRCLE \space v66.0 \\\Circle \space  v69.0 \end{tabular} \\ \cline{2-15}
    
    &  \begin{tabular}[c]{@{}c@{}}PerformanceTiming2\\~\cite{karami2021awakening}\end{tabular} 
    &  &  &  & \checkmark   &  & & \checkmark  &
    & \begin{tabular}[c]{@{}c@{}}\CIRCLE \space v79.0 \\ \XBox \end{tabular} & \begin{tabular}[c]{@{}c@{}}\CIRCLE \space v72.0 \\ \XBox \end{tabular} 
    &  \begin{tabular}[c]{@{}c@{}} \CIRCLE \space v79.0 \\ \XBox \end{tabular}  & \begin{tabular}[c]{@{}c@{}}\CIRCLE \space v12.1 \\ \XBox \end{tabular}  
    &  \begin{tabular}[c]{@{}c@{}} \CIRCLE \space v66.0 \\ \XBox \end{tabular} \\ \hline

    & \begin{tabular}[c]{@{}c@{}}XSS\\~\cite{Chinprutthiwong2020}\end{tabular}
    &  &  &  &  &    & \checkmark  &  &  \checkmark  
    & \XBox & \XBox  & \XBox & \XBox & \XBox \\ \cline{2-15} 
    
    &\begin{tabular}[c]{@{}c@{}} ExtensionHijack\\ {[New]} \end{tabular}&  &  &  &   &  &  
    & &  \checkmark & \davidsstar &\begin{tabular}[c]{@{}c@{}} \CIRCLE \space v82.0\\ \XBox \end{tabular} & \davidsstar  & \davidsstar & \davidsstar  \\ \cline{2-15} 
    
\multirow{-6}{*}{\begin{tabular}[c]{@{}c@{}}Hijacking\end{tabular}} 
   
    & \begin{tabular}[c]{@{}c@{}}LibraryHijack\\ {[New]} \end{tabular}&  &  &  &  &  & \checkmark &  &   \checkmark
    & \XBox  & \XBox & \XBox &  \XBox & \XBox \\ \hline

\bottomrule

\end{tabular}
}
\end{table*}

%% file: sw_mitigations.tex
\section{Existing Mitigations}
\label{sec:mitigations}

In this section, we discuss existing mitigations to some of the attacks
discussed in Section~\ref{sec:attacks}. As mentioned earlier, we have
(re-)produced a proof-of-concept version of the attacks. We then tested the
attacks using multiple versions of five different browsers from different
popular vendors, namely Chrome, Firefox, Edge, Safari and Opera. To make testing
with multiple combinations of browser version and operating systems easier, we
made use of \url{BrowserStack.com}. Our purpose was to estimate {\em when} (at
what browser version) an attack was fixed, whether it was {\em fully or
partially mitigated}, or if the attack is {\em still feasible} in some or all browsers.

\begin{figure}
\setlength{\belowcaptionskip}{-10pt} 
\setlength{\footskip}{30pt}
\setlength{\abovecaptionskip}{5pt plus 3pt minus 2pt} 
    \includegraphics[width=\linewidth]{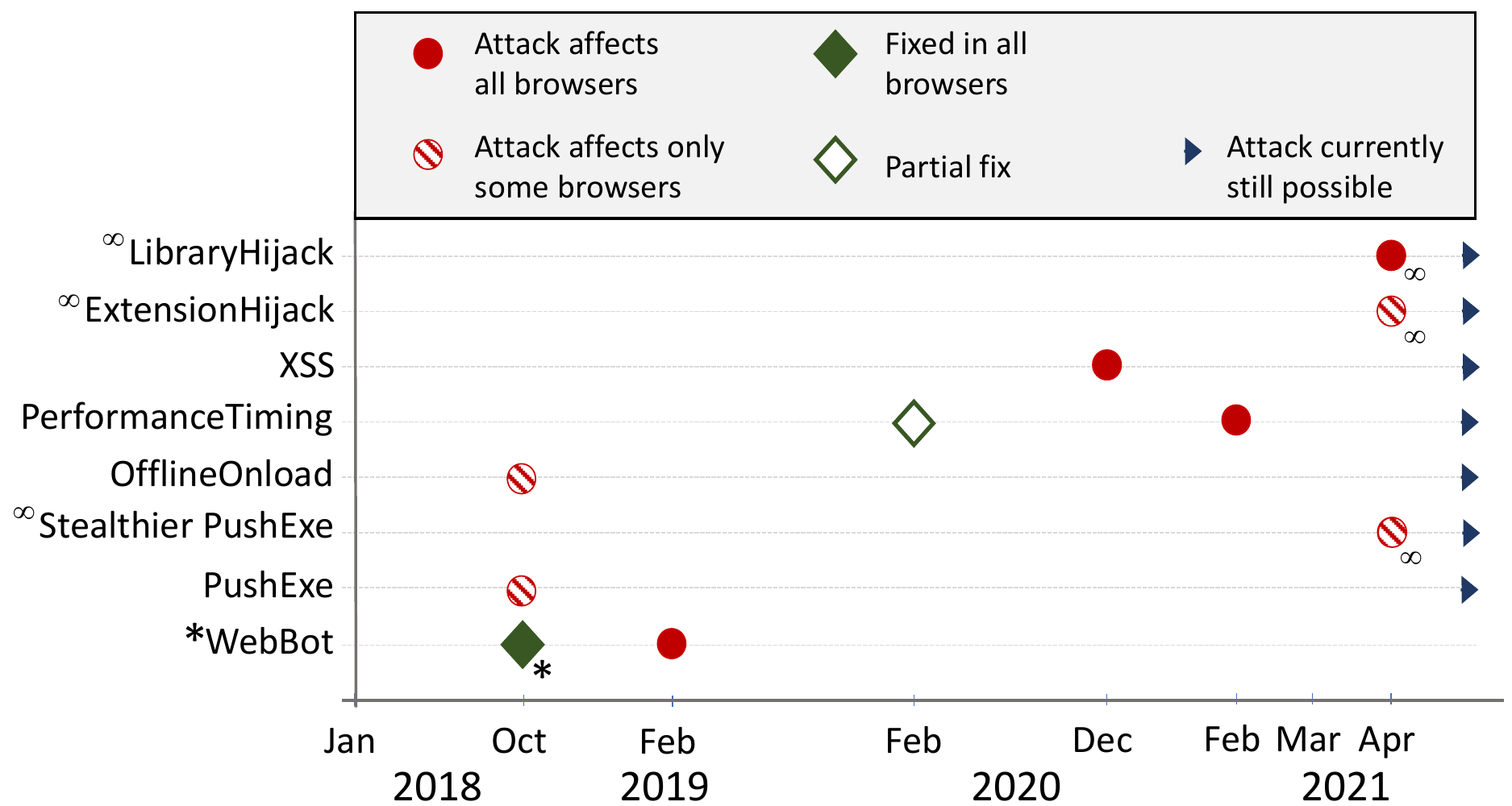}
    \caption{Approximate timeline of attacks publication and mitigations. The $\infty$ symbol denotes new attacks discovered in this paper, which are not yet mitigated. Notice that some mitigations were implemented before official attack publication, perhaps thanks to responsible disclosure (the `*' next to WebBot denotes that a mitigation was described in online documentation related to Firefox, but our own tests show the attack appears to still be possible on that browser).}
    \label{fig:attack_timeline}
\end{figure}

Table~\ref{tab:sok_attacks} provides an overview of browser versions that are
vulnerable to the attacks, and what browser version (if any) fixed or mitigated
the vulnerability. Furthermore, Figure~\ref{fig:attack_timeline} visualizes an
approximate timeline of when the attack was made public and when a mitigation
for the attack was released (if any). 
Overall, we found that some of the attacks have been mitigated by some browsers,
but also that most of the attacks are still possible on at least some of the latest
browser versions. Furthermore, some of the attacks introduced in
Section~\ref{sec:attacks} have not yet been considered for mitigation. Notice
also from Figure~\ref{fig:attack_timeline} that some attacks appear to have been
mitigated by some browsers before the attack was officially published, perhaps
as a result of responsible disclosure processes. Below we discuss what
mitigations have already been implemented or planned so far by browsers, whereas
in Section~\ref{sec:open_problems} we discuss open problems and propose new
mitigations.

\subsection{Mitigating Continuous Execution}
\label{sec:mitigations_contexe}

The continuous execution attacks described in Section~\ref{sec:cont_exe} all
require the victim to visit a malicious website that will register an
attacker-controlled SW on the victim's browser. By design, to avoid overwhelming
the user with UI alerts for SW registration requests, the action of registering
a SW is silently allowed by default and the user is not informed by the browser
that the site she visited registered a SW. The mitigations discussed below still
assume that SWs can be silently registered, but have the effect of limiting the
number of users granting notification permissions, limiting the frequency with which a service worker is activated, or making SW execution
due to web push events less stealthy.

\vspace{3pt}
\noindent \textbf{Termination Delay Limits}. This existing mitigation has the
effect of fully preventing the {\bf WebBot} attack. 
%
%
As explained in Section~\ref{sec:cont_exe}, the {\em WebBot} attack exploited a
self-update behavior by continuously invoking the {\em Update} API, causing the
SW to self-update and continue executing malicious code. To defend against this
vulnerability, major browsers such as Chrome, Edge and Opera have implemented a
limit of up to three minutes on the SW termination timeout (e.g., {\tt
kMaxSelfUpdateDelay} is set to three minutes in the Chromium source code), when
{\em update} is invoked while handling an {\em activate} or {\em install}
event~\cite{sync_issue}. In case of Firefox, according to an online bug report~\cite{update_issue_firefox} a fix was implemented in v70.0. However, while testing our reproduced attack code we were able to keep the SW running for hours. On the other hand, Safari terminates the SW as soon as the related website is closed and is therefore not affected by this attack.
%
%
According to~\cite{sync_issue}, this fix was implemented before the {\bf WebBot}
attack was officially published in~\cite{masterofweb}, as reflected in Figure~\ref{fig:attack_timeline}.

\vspace{3pt}
\noindent \textbf{Notification UI Changes}. Bilogrevic et
al.~\cite{google_usenix} recently showed that although 74\% of all browser
permission prompts that users receive are about notification permissions, only
10\% of these requests are granted by users on desktop devices (21\% on mobile
devices). Because notification permission requests cause unwanted interruptions
during normal user browsing, a new and more quiet notification prompt has been
introduced in Chrome (starting from version 80) and in Firefox~\cite{quiet_ui,
google_usenix, quiet_ui_firefox}. In Chrome, the quiet notification permission prompt is shown within the URL bar when either of two conditions are met: (i) the website
requesting notification permission has a high average deny rate
across its visitors, or (ii) the user recently denied notifications multiple
times (e.g., 3 consecutive time) on different websites within a given
timeframe (e.g., 28 days).  

While we did not find indications that the quiet notification implemented by
Chrome was intended to mitigate specific types of SW abuse, it may have some
mitigating effect on the \textbf{PushExe} and \textbf{StealthierPushExe} attacks
described in Section~\ref{sec:cont_exe}. Because these attacks require victims
to grant notification permission to the attacker's website, the Chrome UI change
may cause the attacker's website to be selected for the quiet notification
prompt, potentially decreasing the number of users that will actually grant the
permission, thus mitigating the attacks. However, as it is not a direct
mitigation to those two attacks, we do not consider the quiet notifications
UI as a fix for the purpose of Table~\ref{tab:sok_attacks}. Furthermore,
in Section~\ref{sec:open_problems} we also discuss how this mitigation may be
easily circumvented by the attacker.

\vspace{3pt}
\noindent \textbf{Default Notifications}. 
%
%
We have confirmed that, at
the time of writing, the {\em PushExe} attack described in
Section~\ref{sec:cont_exe} still works in the latest versions of Firefox
(v91.0).
In Chrome, the {\em PushExe} is not stealthy, because a default notification message\footnote{Message: ``The site has been updated in the background.''} is shown after Chrome detects that no notification is explicitly shown by the SW.
However, we verified that the {\em StealthierPushExe} attack that we introduced in
Section~\ref{sec:cont_exe} remains unmitigated in the latest versions of Chrome (v93.0), Edge (v93.0) and Opera (v78.0).


\subsection{Mitigating Side-Channels}

\noindent \textbf{Event Signaling}. To mitigate the {\bf OfflineOnload} attack
mentioned in Section~\ref{sec:side-channels}, Chrome (at least since v50)
ensures that the {\tt iframe} {\em onload} event is triggered regardless of the
presence of a SW.
However, we were able to verify that this attack is still possible even for the
latest version of Firefox (v91.0) and Safari (v14.0).

\vspace{3pt}
\noindent \textbf{Site Isolation}. The {\bf PerformanceTiming1} attack
summarized in Section~\ref{sec:side-channels} can be mitigated by making sure
that meta-data related to a given origin is not revealed to third-party {\tt
iframe}s. This has been fixed in Chrome since version 83.0. However, as
mentioned in Section~\ref{sec:attacks}, we found that a variant of this attack
appears to be still possible in the latest Firefox browser (v91.0) by monitoring
the {\em initiator} property from a third-party {\tt iframe}. 
Also, our tests with reproduced attack code for the {\bf
PerformanceTiming2} attack confirmed that it still remains unmitigated in the
latest versions of all major browsers (see Table~\ref{tab:sok_attacks}). 

\subsection{Mitigating Other Attacks}
We are not aware of specific mitigations that have already been implemented by
affected browsers to counter hijacking attacks (Section~\ref{sec:hijacking}) or
social engineering attacks (Section~\ref{sec:social_eng}). We will discuss open
problems and potential new mitigations to some of these attacks in
Section~\ref{sec:open_problems}.

%% file: open_problems_mitigations.tex
\section{Open Problems and New Mitigations}
\label{sec:open_problems}

In this section, we revisit some of the attacks presented in
Section~\ref{sec:attacks} and highlight open problems that, to the best of our
knowledge, have not yet been addressed by browsers. Furthermore, we also
propose new mitigations that we believe should be implemented in future browser
versions to address the problems we identified. 

\subsection{Limiting SW Execution}
\label{sec:op_limiting_exe}

\noindent \textbf{Open Problem: }
In Section~\ref{sec:cont_exe}, we discussed different ways (both previously
known as well as new ones) to (silently) extend the execution time of SWs, to
approximately achieve continuous execution. Although some mitigations specific
to the attacks in Section~\ref{sec:cont_exe} have been employed by some
browsers, it may still be possible to create similar attack conditions that
exploit existing or future SW features.
For instance, to circumvent existing mitigations related to always showing
notification messages to users every time a push event occurs (see
Section~\ref{sec:mitigations_contexe}), the SW code could be activated only at a
time when the user may not be paying attention to the screen (e.g., many users
leave the browser always open, even at night), as also discussed
in~\cite{pridepwa}. Furthermore, even if the SW is activated a large number of
times in a row using many consecutive push messages, the SW can prevent the
browser from showing multiple notifications from the same website, which may
make the user suspicious. To make sure that the user will only see one single
notification, the SW can keep reusing the same {\tt tag} parameter value, as
shown in Listing~\ref{code:notification_reuse} in Appendix. Fundamentally, we
found that browsers do not currently put any constraints on the number of push
messages a SW can receive or on the amount of execution time granted to any
given SW. This leaves open possible abuse paths, as exemplified above.

\vspace{3pt}
\noindent \textbf{Proposed Mitigation: } 
To defend against present and future {\em continuous execution} attacks, we need
a more generalized defense that can dynamically monitor the SW execution time
and throttle it when abuse is suspected. This can be accomplish with additional browser policies. Specifically: 

\begin{enumerate}
    \item Monitor and limit the overall background execution time for which a SW
    runs every time it is activated. This will prevent known and unknown ways of
    artificially elongating the time a SW remains active (e.g., this would
    mitigate abuse vectors similar to the self-update exploit used in the {\bf
    WebBot} attack).
    \item Limit the number of push events received by a SW within a predefined
    time window. This would have the effect of limiting the frequency with which
    a SW can be remotely activated, thus throttling continuous execution
    attacks.
    \item Ensure that a SW notification displayed to the user remains visible
    until the user interacts with it (e.g., by clicking on it or closing it
    explicitly). This would help to mitigate stealthy activations via push
    events.
    \item Limit the volume of third-party network requests issued in the
    background by a SW. While this is not strictly a limitation on execution
    time, it can be useful to mitigate possible ``bursty'' bandwidth-exhaustion
    DDoS attacks (e.g., by issuing many background network requests in a short
    execution time) against third-party websites.
\end{enumerate}

In Section~\ref{sec:policy_impl}, we discuss how we implemented a
proof-of-concept version of some of these policies in Chromium. In Section~\ref{sec:measurements}, we measure how SWs are currently used by popular
websites and propose concrete thresholds to limit SW execution with limited or no impact on legitimate SW functionalities.

\subsection{Limiting Malicious SW Permissions}
\label{sec:limiting_permissions}

Because by design SWs can be silently registered by any website, preventing the
registration of an arbitrary SW may not be possible\footnote{Obviously, blocking
a known malicious website, for instance by using URL blocklists, would also
prevent any related SW to be registered. Unfortunately, threat feeds and
blacklists often have gaps and may not block a malicious site for a prolonged
time~\cite{Bouwman2020}, during which many victims could visit the SW and
have a SW installed.}. However, notice that without being granted the
notification permission the SW cannot receive push messages and the attacker is
unable to launch effective {\em continuous execution}  attacks
(Section~\ref{sec:cont_exe}) or {\em social engineering} attacks
(Section~\ref{sec:social_eng}), thus limiting the damage that a malicious SW may
cause. 
The new quiet notification permission requests described in~\cite{quiet_ui} (see
also Section~\ref{sec:mitigations_contexe}) could therefore be seen as a way of greatly
restricting the damage a malicious SW can do. The reason is that, presumably,
only few users would grant notification permission to an untrusted website
(notice also that the permission grant rate is already low in general for most
websites~\cite{google_usenix}). Thus, it is likely that a malicious site that
asks its visitors for notification permission would rapidly meet the criteria to
qualify for the quiet notification UI. In turn, this may have the effect of
further reducing the number of users who grant permission and whose browser can
be meaningfully abused by the malicious SW.

\noindent \textbf{Open Problem: }
Unfortunately, the quiet notifications UI is not in itself an effective
mitigation for limiting the number of victims that may grant notification
permission to a malicious SW. One reason for this is that malicious SWs can
leverage the same {\em double permission} prompt
pattern~\cite{double_permission_chrome, double_permission_onesignal} that is
recommended as a good practice to legitimate web developers. The {\em double
permission} prompt consists in asking the user twice whether they would like to
receive notifications from a website. The first time, the website uses
JavaScript code~\cite{double_permission_chrome, double_permission_onesignal} to
create a notification permission dialog box within the page's context (see
Figure~\ref{fig:double_perm} in Appendix). Only if the user confirms, typically
by clicking on a custom ``Yes'' or ``Sign up'' button, the SW will go ahead an
request the actual notification permission through the browser UI. The reason
why legitimate websites often use this pattern is because they want to avoid
being blocked from asking the user for notification permission again in the near
future. Since the website controls the JavaScript dialog box, the browser will
not be aware that the user may want to block notifications from this site, and
therefore the website gets to ask again every time the user visits it. The net
effect of this legitimate (and recommended) web development pattern is that the
browser may grossly overestimate the notifications allow-rate for a given
website. Intuitively, it is highly likely that users who do not want to receive
notifications from a website will click on the ``Not now'' or ``Dismiss'' (or
equivalent) button on the JavaScript dialog box and they will not be presented
with the real SW's request for notification through the browser. Ultimately,
given that one of the criteria to enable quiet notifications UI is a higher
denial rate, malicious sites can evade this by simply adopting the double
permission pattern. In general, the very recommendation to legitimate web
developers on adopting the {\em double permission} prompt may make the newly
introduced quiet notification UI much less effective than anticipated.
%

Another issue is due to the fact that the attacker's site could
also launch social engineering attacks similar to the ones mentioned
in~\cite{phani_imc} to encourage the user to explicitly click on the quiet UI's
icon and explicitly grant permission.

\vspace{3pt}
\noindent \textbf{Proposed Mitigation: }
Unfortunately, preventing users from granting notification permission to a
malicious SW may be difficult, as discussed above. In addition, once a SW is
registered and has been granted permission, the SW will persist until the user
explicitly removes them following a cumbersome process that involves going
through the browser preferences and settings. If little or no constraints are
imposed, this may lead to significant abuse, as described in
Section~\ref{sec:attacks}.

As a mitigation, we argue that the browser should monitor each SW's behavior for
signs of abuse. The browser could then explicitly offer the user to de-register
a SW (with a specific UI dialog box) when anomalous behavior is detected, or it
could automatically de-register the SW.

For instance, consider the following scenario:
\begin{itemize}
    \item The user visits a malicious website once, at which point a SW is
    registered and notification permission is requested.
    \item Assume that the user grants notification permission at first visit
    (e.g., due to a social engineering attack), and that the user never visits
    the site again.
    \item Afterwards, the SW runs frequently in the background (e.g., due to
    frequent push events) to achieve continuous execution using one
    of the approaches described earlier.
\end{itemize}

In this example scenario, the browser could detect that the SW is violating one
or more of the policies we proposed in Section~\ref{sec:op_limiting_exe}, which
will automatically limit SW execution. At the same time, the browser could
detect that the user has not visited the site again since the first time the SW
was registered, or that the site has a very low engagement score as defined by
Chromium\cite{site_engagement}. In this case, the browser could inform the user
and ask whether she would like to de-register the SW. To make the decision
easier, the browser could let the user know that the SW has been running
anomalously and frequently, and that the user has visited the website only once
(or very rarely). As an alternative, if the SW is not explicitly de-registered
by the user and the browser continues to observe that the SW abuses execution
limit policies, it may simply de-register the SW outright (notice that the SW
could always be re-registered next time the user visits the same site, if the
user so desires).

\subsection{Restricting Third-Party Code Inclusions}
\label{sec:op_csp}

\noindent \textbf{Open Problem: }
It is well known that third-party JavaScript code inclusions come with security
risks~\cite{Nikiforakis2012}. As discussed in Section~\ref{sec:hijacking}, the
common practice of including third-party code into SWs could lead to {\em
hijacking} attacks. Content Security Policies~\cite{sw_policies} (CSPs) can be
used to defend against SW {\em hijacking attacks} such as {\bf XSS} or {\bf
LibraryHijack}, for instance by using the {\tt script-src} to restrict imported
code into a SW to be loaded only from authorized domains. 
However, implementing this defense is up to web developers, and in
Section~\ref{sec:measurements} we show that only a small fraction of websites
express SW-specific CSP restrictions (also, low CSP adoption is a known
issue in general~\cite{Weissbacher2014}). Unfortunately, when
CSP policies are missing, the browser poses no restrictions to importing
third-party code into a SW.

\vspace{3pt}
\noindent \textbf{Proposed Mitigation: }
We argue that, given the potential for abuse related to SWs, the browser should
follow the {\em fail-safe defaults} principle and deny the ability to import
third-party code by default. Namely, the browser should always assume a default
{\tt script-src: `self'} policy for SWs. The web developer could then express
exceptions to this default policy by explicitly listing authorized third-party
origins in the {\tt script-src} directive (this CSP directive would need to be
sent to the browser with every SW file response, which can be easily configured
in modern web servers). In Section~\ref{sec:measurements}, we will also show that the number of different origins that would need to be authorized in current production SWs is very small (only one or two, if any).

Unfortunately, {\tt script-src: `self'} does not
prevent {\tt eval()} to be used in SW code~\cite{eval_worker}, leaving a door
open to potential code hijacking attacks such as variants of the {\bf XSS}
attack proposed in~\cite{Chinprutthiwong2020}. Instead, the use of {\tt eval()}
should be disabled by default and enabled explicitly by adding the directive
{\tt script-src: 'unsafe-eval'}, as for page JavaScript code.
%
Notice also that the {\tt worker-src} CSP directive~\cite{worker_src} 
can be used to restrict what URLs may be used to load a SW file, but does not
apply to the {\em importScripts} API. Furthermore, {\tt worker-src} does not
have any effect on blocking the use of {\tt eval()} either. 

\subsection{Restricting the Scope of Third-Party Libraries}
\label{sec:restricting_tpc}

\noindent \textbf{Open Problem: } 
In some cases, web developers may want to explicitly allow third-party services,
such as {\em push services}, to include code into their SWs. For instance,
assume that website $W$ wants to make use of push service $P$ (e.g.,
OneSignal.com, iZooto.com, etc.). In this case, $W$ would want to import $P$'s
third-party code into its SW, $S_W$ (notice that $P$'s origin can be
easily specified in $S_W$'s {\tt script-src} CSP directives, as discussed
earlier). Unfortunately, once $P$'s code is imported in the context of $S_W$,
there is no way to restrict what APIs $P$'s code can use, thus potentially
enabling a {\bf LibraryHijack} attack (see Section~\ref{sec:hijacking}).

To attempt to isolate their third-party code from the first-party website's
$S_W$, web developers could register a separate SW, $S_P$, with a different {\em
scope}, instead of running the SW under the root path of $W$. This would allow
$S_P$ to coexist with other SWs registered under $W$. In addition, $S_P$ would
not be able to intercept network requests related to content outside of its
scope, thus effectively isolating the third-party SW code. However, while this
would be an improvement, it does not prevent $S_P$ from being able to directly
accessing $W$'s cookie store~\cite{cookie_store}. Furthermore, $S_P$ would also
be able to access the cache\cite{cache_api} and thus any content previously
stored by $S_W$, since there is currently no cache isolation for SWs registered
with different scopes under the same origin. Consequently, the third-party code
could still potentially access highly sensitive information related to $W$.

\begin{lstlisting}[caption={Proposed change to register SW with limited capabilities.},captionpos=b,language=JavaScript, label=code:sw_options]
    if ('serviceWorker' in navigator) {
      // proposed register() options to sonly enable use of push notifications // while prevent the use of other sensitive APIs    // such as cookie store, cache, fetch events, etc.
      navigator.serviceWorker.register('/pushservice_sw.js', 
            {scope: './pushservice_sw_scope/', 
             capabilities: 'push', 'notifications'})
     }
\end{lstlisting}

\vspace{3pt}
\noindent \textbf{Proposed Mitigation: } \karthika{To mitigate the risk of {\bf
LibraryHijack} attacks, a possible approach is to explicitly limit the {\em capabilities} that a given SW script can have. This list of capabilities could be expressed at registration time.
At the moment, when registering a SW under a given origin (via {\tt
ServiceWorkerContainer.register()}~\cite{sw_reg}), only the {\em scope} of
the service worker can be limited. However, we argue that the {\tt options}
parameter should be extended to allow expressing additional constraints. For instance, we could express what set of functionalities or APIs the SW is allowed to access, or what set of events it is allowed to listen to. This way, we could restrict the capabilities of a third-party SW (i.e., a SW that imports third-party code) to using the {\em push} and {\em notifications} APIs while denying the use of the cookie store, the cache, or the fetch API, as in the example code in Listing~\ref{code:sw_options}. On the other, first-party SW scripts could be registered without capability restrictions so that they can use any functionality made available to SWs by the browser}.

\karthika{More fine-grained changes would also be useful, such as expressing whether the SW is allowed to access the cache but at the same time indicate whether the cache for this SW should be isolated by scope (notice that this proposed cache isolation mechanism could be implemented in a way similar to the cache isolation approach used for the now-deprecated {\em AppCache} API~\cite{appcache_api}).
These browser modifications would
still allow push services to provide a convenient way of managing push
notification campaigns on behalf of a website $W$ while limiting exposure of
potentially sensitive information belonging to $W$'s users. Notice also that the
proposed fine-grained SW policies would be somewhat analogous to {\em
Feature Policy} for {\tt
iframes}~\cite{feature_policy}.}

\subsection{Restricting Extensions Permissions}

\noindent \textbf{Open Problem: }
Even if the mitigations described in earlier sections are implemented, a website's SW could still be hijacked by extensions, using the {\bf ExtensionHijack} attack described in Section~\ref{sec:hijacking}. Furthermore, extensions can tamper with CSP directives in network responses~\cite{pantelaios2020you}, which can further facilitate SW hijacking attacks.

\noindent \textbf{Proposed Mitigation: }
Other major browsers, such as Firefox, should follow Chromium's approach and
prevent extensions from interfering with responses to SW file requests (e.g., by
preventing access to the {\em FilterResponse} API).




%% file: evaluation.tex
\section{Measuring In-The-Wild SW Behavior}
\label{sec:measurements}

In this section, our objective is to measure the behavior of in-the-wild SW code
used by popular websites. The main goal is to learn how production SW code may
be impacted by some of the mitigations we discussed in
Section~\ref{sec:open_problems}. We focus mostly on mitigations against
coninuous execution attacks (Sections~\ref{sec:op_limiting_exe}
and~\ref{sec:limiting_permissions}) and potentially malicious third-party code
inclusions (Sections~\ref{sec:op_csp} and~\ref{sec:restricting_tpc}), and aim to
learn how policy enforcement thresholds may be tuned to curtail possible attacks
while minimizing their impact on legitimate SW behaviors.

\subsection{Browser Instrumentation for SW Forensics}
\label{sec:instrumentation}

To obtain fine-grained information on the behavior of SW code for real-world web
applications, we first developed an instrumented version of the Chromium browser
(v84.0.4147.121) with an embedded {\em Service Worker Forensics} engine. 
%
%
%
Our SW forensics engine logs fine-grained information
regarding the following:

\begin{itemize}
    \item The occurrence of SW life-cycle events such as {\em install}, {\em
    activate}, {\em update}, {\em uninstall}, and {\em termination}.
    \item Any permissions that were requested by the SW code, such as push
    notifications and geo-location. In addition, we automatically grant these
    permissions to monitor how their related APIs are utilized.
    \item Detailed information about network requests issued by SWs, including
    the URL of resources being fetched.
    \item API calls made by SW code with respect to caching, push, and
    notification APIs.
    \item CPU and memory consumption, and network usage (e.g., number of
    third-party network requests and related URLs) for each SW instance.
    \item We also simulate user interactions with the browser that are required
    to trigger events to be handled by a SW.
\end{itemize}

The logs generated by our forensics engine are then analyzed offline to
measure useful properties about the behavior of SWs in the wild. Since Chromium
serves as a basis for many popular browsers, such as Chrome, Opera, Edge, etc.,
the measurement results we obtained can be considered as representative of SW
code running on a variety of browsers.

\subsection{Experimental Setup}

Because the main objective of our measurements is to understand how real-world
SW code behaves with respect to the mitigations we proposed in
Section~\ref{sec:open_problems}, we focus on analyzing SWs registered by the
most popular websites according to Alexa.com~\cite{alexa}. We organize our
measurement results by dividing the top Alexa websites into different {\em
bands}, based on their ranking (e.g., 0-1K, 1K-5K, 5-10K, etc.), as shown in
Table~\ref{tab:alexa_Ranking}. Different rank bands provide insights into the
behavior of websites at different levels of popularity. In total, we identified
5,918 websites that registered at least one SW, with 5,309 sites among the top
100k ranking and an additional 609 sites with rank $>$100k. However, notice that
only a subset of these SWs request notifications permission.

Since we are especially interested in mitigations against continuous execution
attacks, we focused our investigation on\karthika{ 1,750 (out of 5,918 websites that
register a SW) whose SW code requested notifications permission, and created a
small farm of automated instrumented browsers (see
Section~\ref{sec:instrumentation}) to interact with these web pages.} For each of
these 1,750 web applications, we continued interacting with them and monitored
their SW behavior for 3 days. To drive our instrumented browser to automatically
visit and interact with these web pages we made use of custom
Puppeteer~\cite{puppeteer} scripts.

\karthika{
Notice that our data collection and analysis does not include websites such as
social media and messaging apps that may require login to send notifications to
users. The reason is that for these websites the behavior of their SWs, such as
the number and frequency of web push notifications, is highly dependent on
user activities and social network. We
exclude these sites from our measurements, as it would be highly
challenging to simulate a realistic network of users that send messages to each
other in a way that is representative of a large and diverse user population. However, it should be noted that SW
security policies that aim to impose limits on push notifications may also
include a customizable allow-list for popular social media and messaging
websites.}

\input{tables/alexa_bands}




\begin{figure*}[t]
\begin{center}
 \begin{subfigure}{.33\textwidth}
    \begin{center}
        \includegraphics[width=0.9\linewidth]{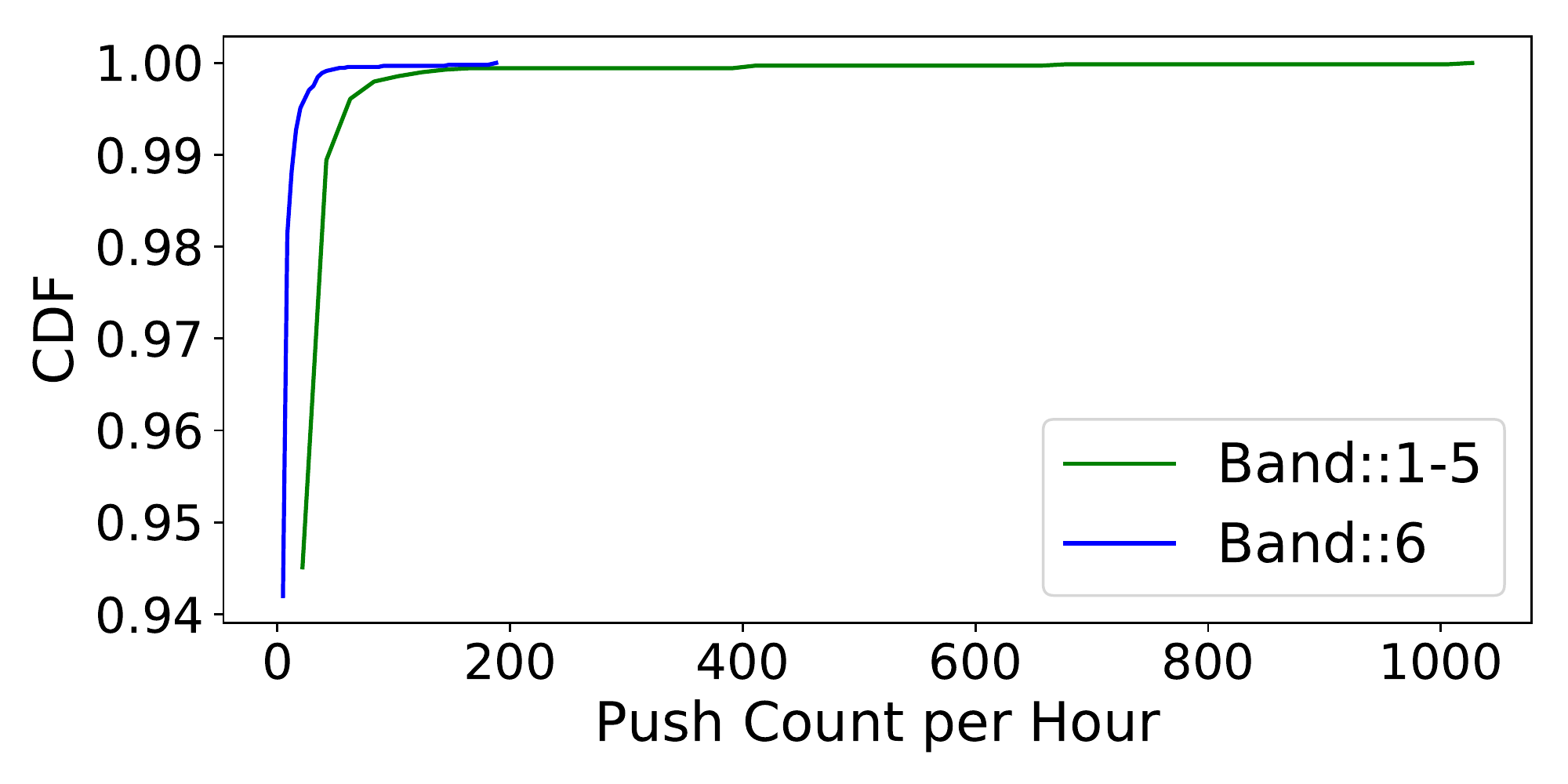}
        \caption{}
        \label{fig:push_counts}
    \end{center}
 \end{subfigure}%
\begin{subfigure}{.33\textwidth}
    \begin{center}
        \includegraphics[width=0.9\linewidth]{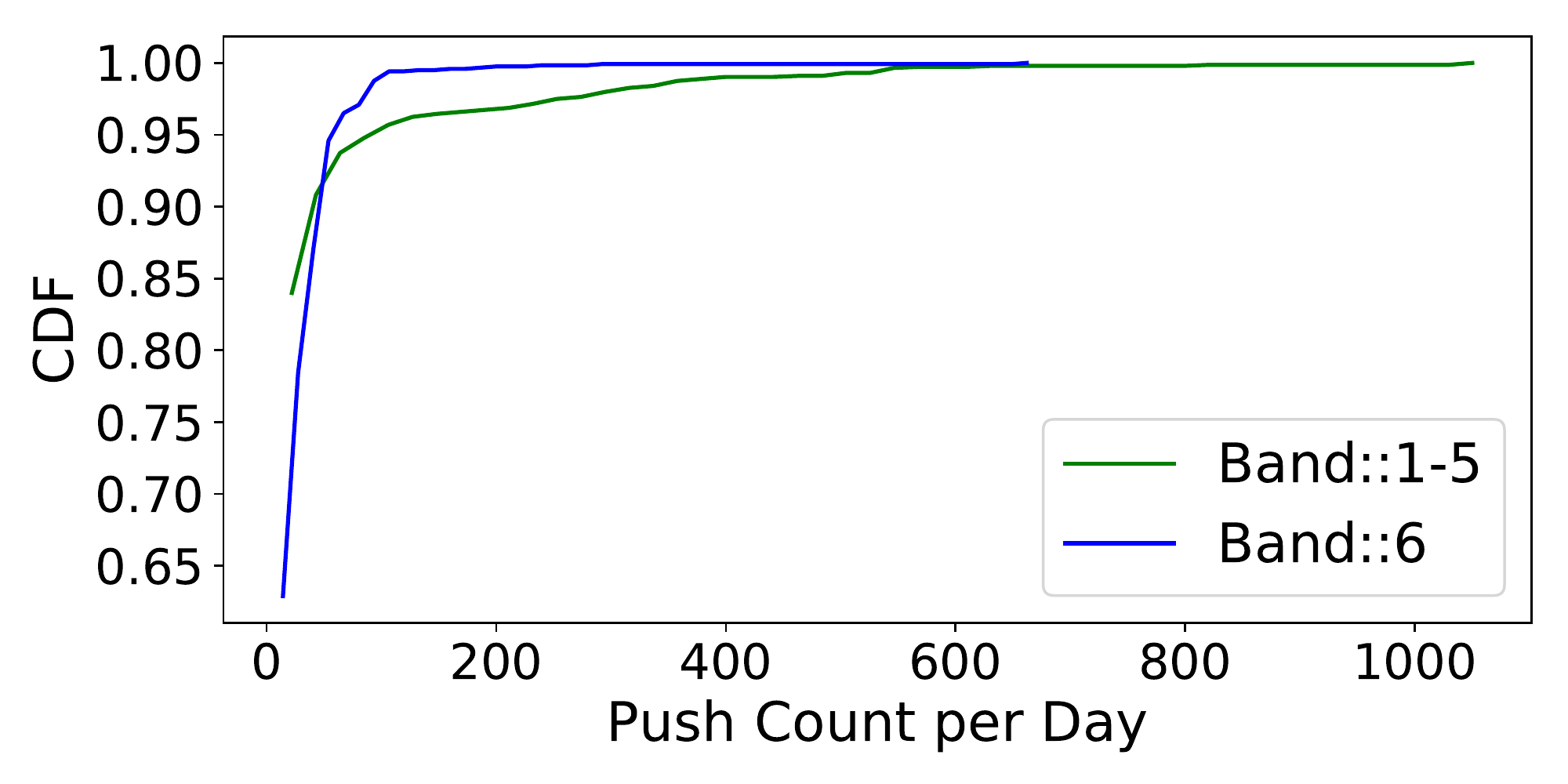}
        \caption{}
        \label{fig:push_counts_day}
    \end{center}
 \end{subfigure}%
 \begin{subfigure}{.33\textwidth}
    \begin{center}
        \includegraphics[width=0.9\linewidth]{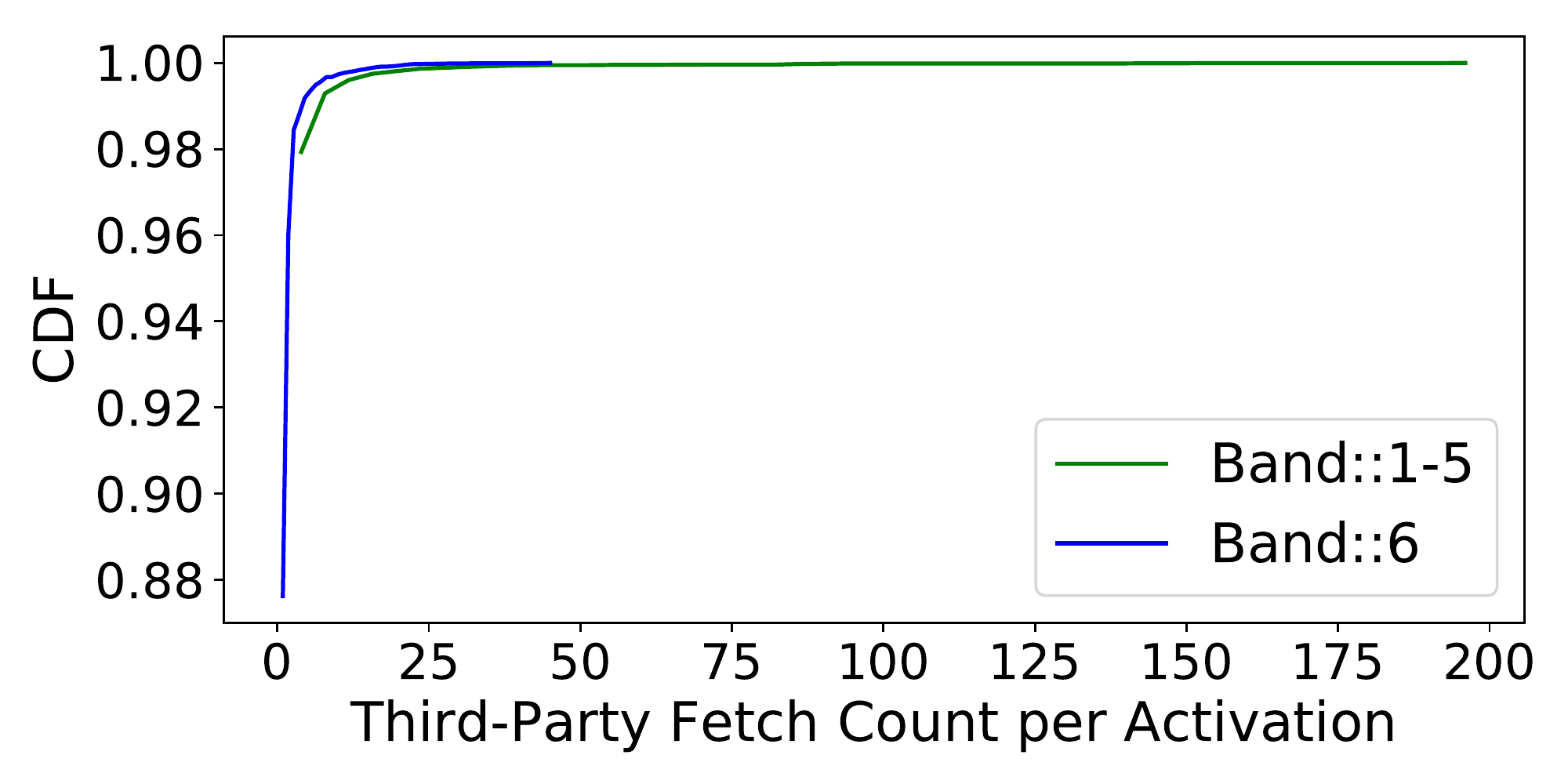}
        \caption{}
        \label{fig:tp_fetch_count}
    \end{center}
 \end{subfigure}\\[2ex]
\begin{subfigure}{.33\textwidth}
    \begin{center}
        \includegraphics[width=0.9\linewidth]{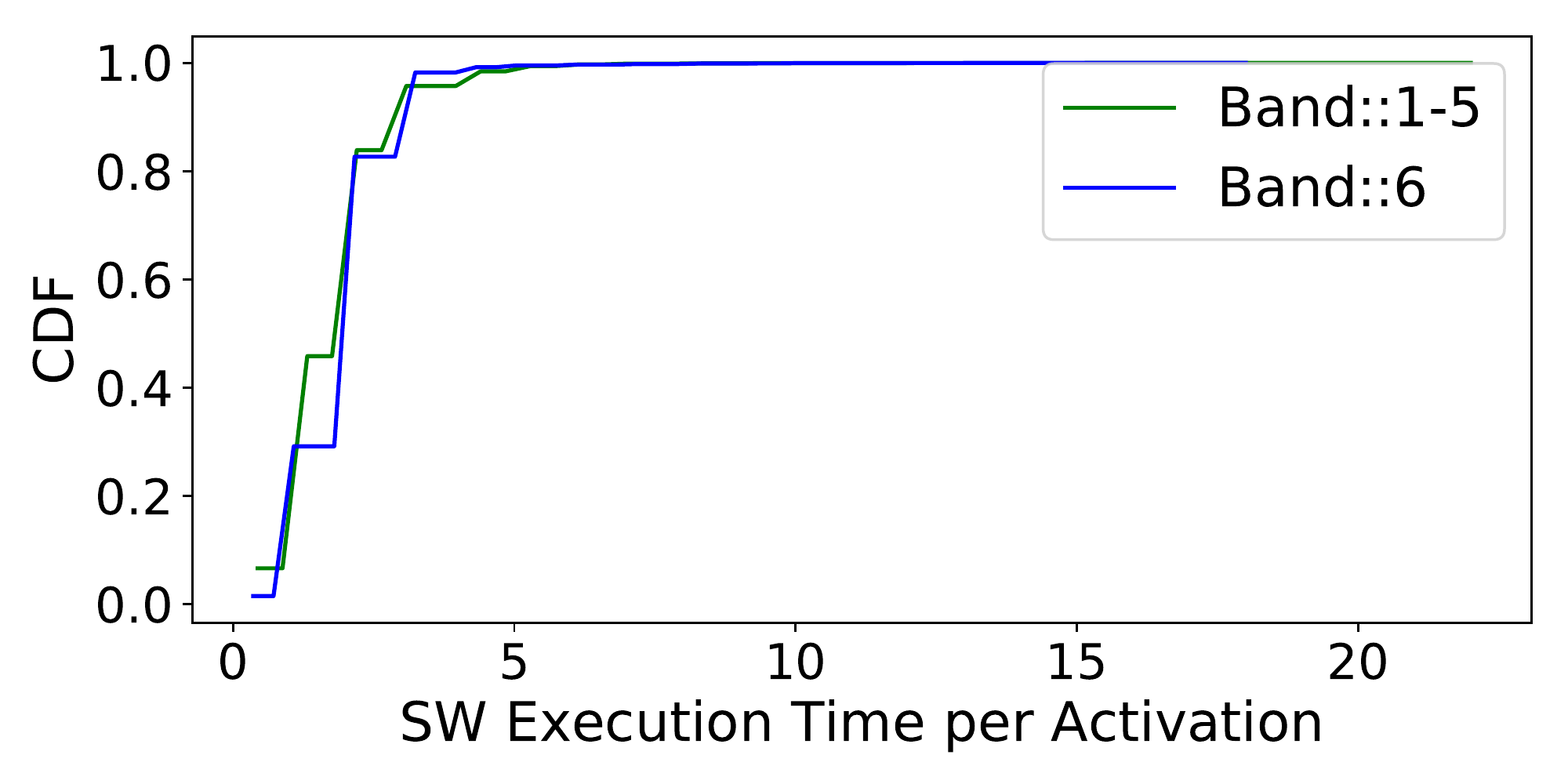}
        \caption{}
         \label{fig:sw_lifetime}
    \end{center}
 \end{subfigure}%
 \begin{subfigure}{.33\textwidth}
    \begin{center}
        \includegraphics[width=0.9\linewidth]{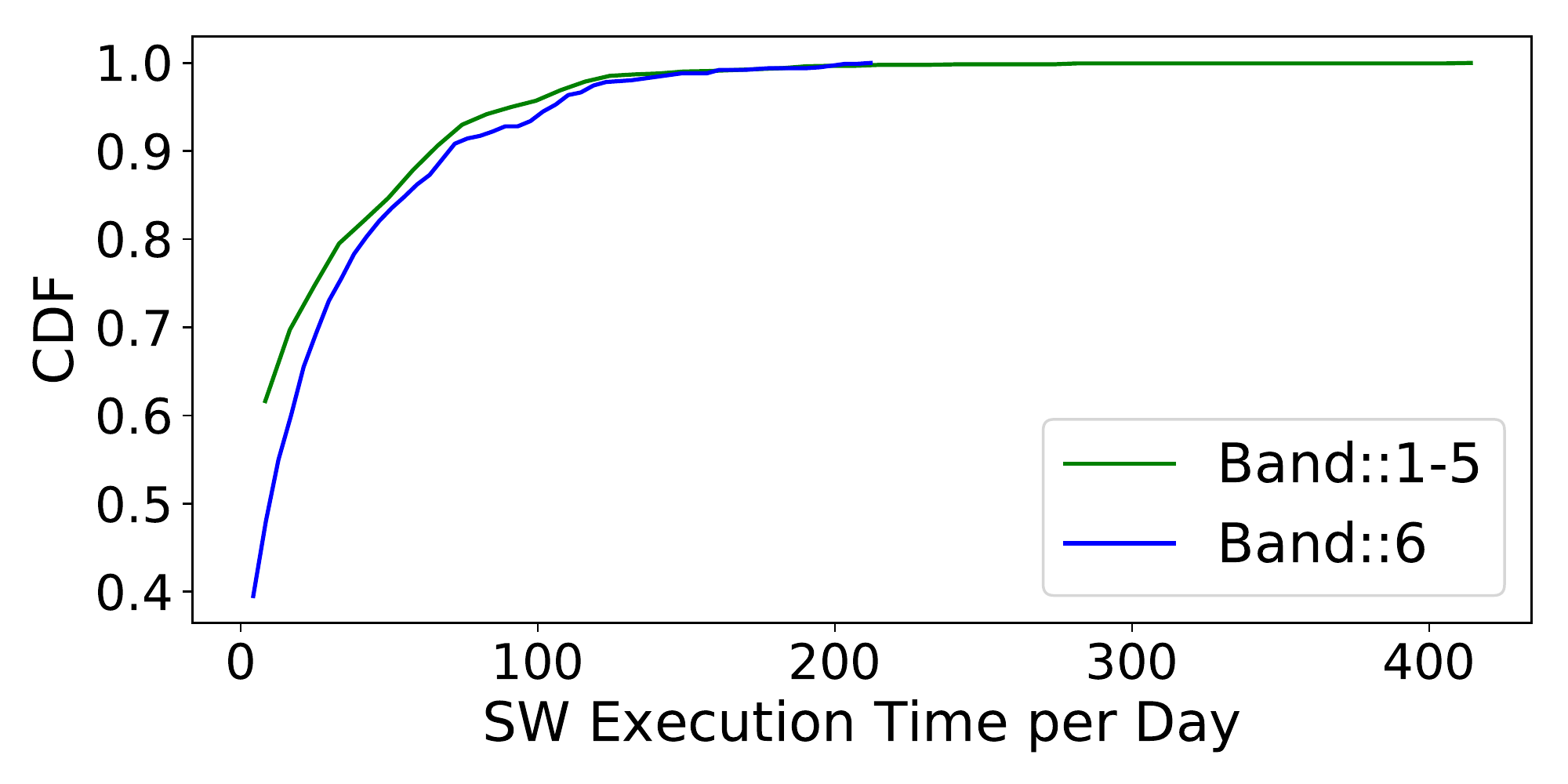}
        \caption{}
         \label{fig:sw_lifetime_per_day}
    \end{center}
 \end{subfigure}\\[2ex]
 
  \caption{SW behavior measurements. Each graph displays the distribution (CDF) of occurrences of an event within a specific time window: a) Push count per hour; b) Push count per day; c) Third-party fetch count per SW activation; d) SW execution time per activation (in minutes); e) SW execution time per day (in minutes).}
  \label{fig:policy_values}
\end{center}
\end{figure*}
\vspace{5pt}

\subsection{SW Behavior Results}

In Section~\ref{sec:op_limiting_exe} we discussed a number of restrictions that
could be imposed on SWs to limit their execution time and reduce potential
damage that a malicious SW may cause due to {\em continuous execution} attacks
(see also Section~\ref{sec:cont_exe}). Specifically, among other mitigations
(see Section~\ref{sec:op_limiting_exe} for details) we proposed to (a) limit
overall background execution time, (b) limit the number of push events within a
given time window, (c) ensure notifications are visible to users, and (d) limit
the volume of third-party network requests.

In the following, we measure how current production SWs behave, compared to the
limitations listed above. This will help us in two ways: (i) determine how
different limits may impact existing SW behaviors, and (ii) inform the choice of
thresholds that could be used in the implementation of new SW security policies.
Detailed measurement results are reported in Figure~\ref{fig:policy_values} and
Table~\ref{tab:percentile_bands}, which we discuss below.

\subsubsection{Frequency of Push Events}
\label{sec:push_limit}
Among the 1,750 websites we monitored, 518 of them have a SW that received at least one push event during our analysis period (i.e., 3 days). To estimate the frequency with which push events are received by our instrumented browser, we divide the timeline into slots of one hour, and count the number of push events per hour for each SW. Figure~\ref{fig:push_counts} shows the distribution (a CDF) of the number of push events per hour across all monitored SWs. Also, from Table~\ref{tab:percentile_bands} we can see that at the 90th percentile, SWs receive 14 push events or less per hour. 

\begin{figure}
    \setlength{\belowcaptionskip}{-15pt} 
    \setlength{\footskip}{30pt}
    \setlength{\abovecaptionskip}{5pt plus 3pt minus 2pt} 
    \centering
    \includegraphics[width=0.8\columnwidth]{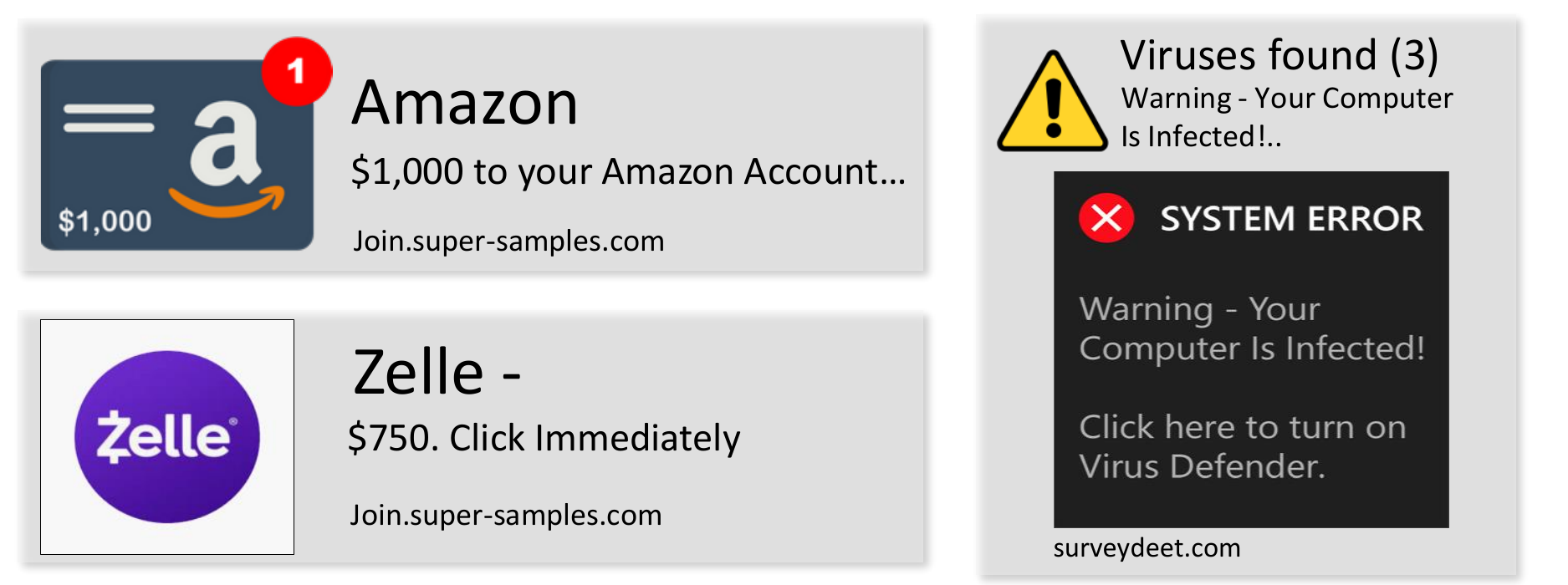}
    \caption{Examples of spam/malicious notifications}
    \label{fig:spam_notifications}
\end{figure}

From Table~\ref{tab:percentile_bands}, we can see that if we implemented a SW
security policy that limits the number of push events per hour to $\leq 14$,
this policy would affect (i.e., throttle) the SWs of 49 different websites (49
is the sum of the number of SWs under each ranking band), with almost half of the
impacted websites having a ranking above 100k (Band-6). While at a first glance
this result may look like a potentially significant impact on production SWs, a
detailed manual analysis of the push notifications that would be curtailed by
the new policy reveals something different. In fact, we found that all 49
websites that would be potentially impacted sent notifications that could be
considered abusive. Specifically, the notifications we recorded from those sites
are related to (potentially malicious) WPN ads, which were previously also
identified by other researchers in~\cite{subramani2020} as being often abusive (some example
notifications reconstructed from our logs are reported in
Figure~\ref{fig:spam_notifications}). Thus, it appears that the proposed limit
on the frequency of push messages would at most throttle the number of
push-based ads received by users, without significantly affecting most
legitimate SWs. At the same time, limiting the number of push events that can
activate a SW can help to decrease the potential for continuous execution
attacks that may be used for instance to perform
cryptomining, DDoS attacks, or other malicious tasks, as further discussed
later.

\subsubsection{Execution Time}
For each SW, we also measured the maximum execution time per activation. Namely,
we measure the time for which a SW ran without releasing control or being forced
to stop by the browser (as before, these measurements were performed throughout
our 3 days of monitoring per each web application's SW code).

As it can be seen in Figure~\ref{fig:sw_lifetime} and
Table~\ref{tab:percentile_bands}, at 99\% of the instances, SWs were alive for a
maximum duration of 5 minutes per activation. At the same time, we also found
that 20 websites had a SW that at some point remained active beyond the maximum
limit (5 minutes) allowed by the browsers. The maximum continuous execution time
per activation that we observed was 22 minutes.
%
%
These cases of long continuous execution were possible because the SW
termination was delayed by the browser as the SW received multiple events (e.g.,
multiple consecutive push events) in close succession, with the next event
arriving and being handled before the SW finished handling the previous event.
This again demonstrates that the possibility of abusing SWs to perform malicious
tasks such as cryptomining and DDoS attacks remains open.

As in Section~\ref{sec:op_limiting_exe}, we argue that the browser should impose
stricter limits to continuous SW execution. For instance, it could be limited to
5 minutes, since 99\% of all SWs activations we measured never exceeded this
threshold. Longer continuous execution times should be considered as anomalous
and potentially dangerous.

Besides the execution time per activation, we also calculate the overall SW execution time per day, as the sum of the execution time spent during all activations of a given SW for a day of observation. As shown in Figure~\ref{fig:sw_lifetime_per_day}, 95\% of SWs were active for less than 90 minutes per day. However, we found 17 websites whose SWs were active from 146 up to 400 minutes (over 6 hours). By analyzing the logs, we found that these websites ``spammed'' the browser with a large number of potentially malicious notifications (similar to Figure~\ref{fig:spam_notifications}).
%
%
As an example of a website whose SW exhibited long execution times, we found that \url{waploaded.com} (ranked 51,299) registered a SW that sent over 50 push events per hour in multiple time windows, and as a result activated its SW and kept it running for long periods (e.g., 22 minutes in one single activation and over 4 hours in a single day).

\input{tables/percentile}

\subsubsection{Third-party Background Network Requests}

To reduce the risk and impact of SWs participating in DDoS attacks, in
Section~\ref{sec:op_limiting_exe} we proposed to limit the volume of third-party
network requests that the SW could issue while in the background (i.e., when the
related web application is not rendering on a browser tab).

To understand what may be a good volume threshold, we measured the number of
fetch requests that were made to third-party origins by each SW while running
{\em in the background}. Specifically, to identify these requests we perform two
different checks: (a) first, we make sure that the request was issued by a SW by
checking whether the JavaScript execution context that issued the fetch request
belongs to a SW script\footnote{As determined by calling {\tt
IsServiceWorkerGlobalScope()} of {\tt ExecutionContext}.}, then (b) we make sure
the request was executed in the background by checking if it was made from
inside a {\em fetch} event listener, which would indicate that it was invoked
when a page request is handled by a SW and thus it was not issued in the
background. To verify (b), we should notice that whenever a {\em fetch} listener
is started, it invokes the {\tt StartFetchEvent} method and at its completion it
invokes {\tt DidHandleFetchEvent} under {\tt ServiceWorkerGlobalScope}. We log
these calls and filter out any fetch requests made between these two events,
since they are not background requests. At the end of this filtering process, we
are left with the {\em background network requests} made by SWs. 

To account for network requests to explicitly authorized third-parties, such as
push services that are intentionally imported into a SW's code, we first
determine the domain name associated with all URLs in {\em importScripts} calls
and exclude them from our background network requests statistics (i.e., network
requests from a SW to its push service domains are effectively counted as
first-party requests). After the filtering explained above, we found that 99\%
of all SWs issued no more than 5 background network requests to third-party
origins per each activation.


Although we did not find any evidence of in-the-wild SWs that performed malicious attacks such as DDoS attacks, we were able to reproduce attack code that can indeed send a large number (e.g., 50 per second) of third-party background network requests with no browser limitations. Therefore, we believe that limiting the number of such background requests (e.g., to $\leq 5$) per activation, in combination with limiting the frequency of activations due to push events, as discussed in Section~\ref{sec:push_limit}, is necessary to significantly reduce the risk for SW-based DDoS attacks while having minimal or no impact on the vast majority of legitimate SWs.



\subsubsection{Third-Party Code Inclusions}

\begin{figure}[!ht]
\setlength{\belowcaptionskip}{-6pt} 
\setlength{\footskip}{30pt}
\setlength{\abovecaptionskip}{5pt plus 3pt minus 2pt} 
    \centering
    \begin{subfigure}{0.48\columnwidth}
    \centering
     \includegraphics[scale=0.25]{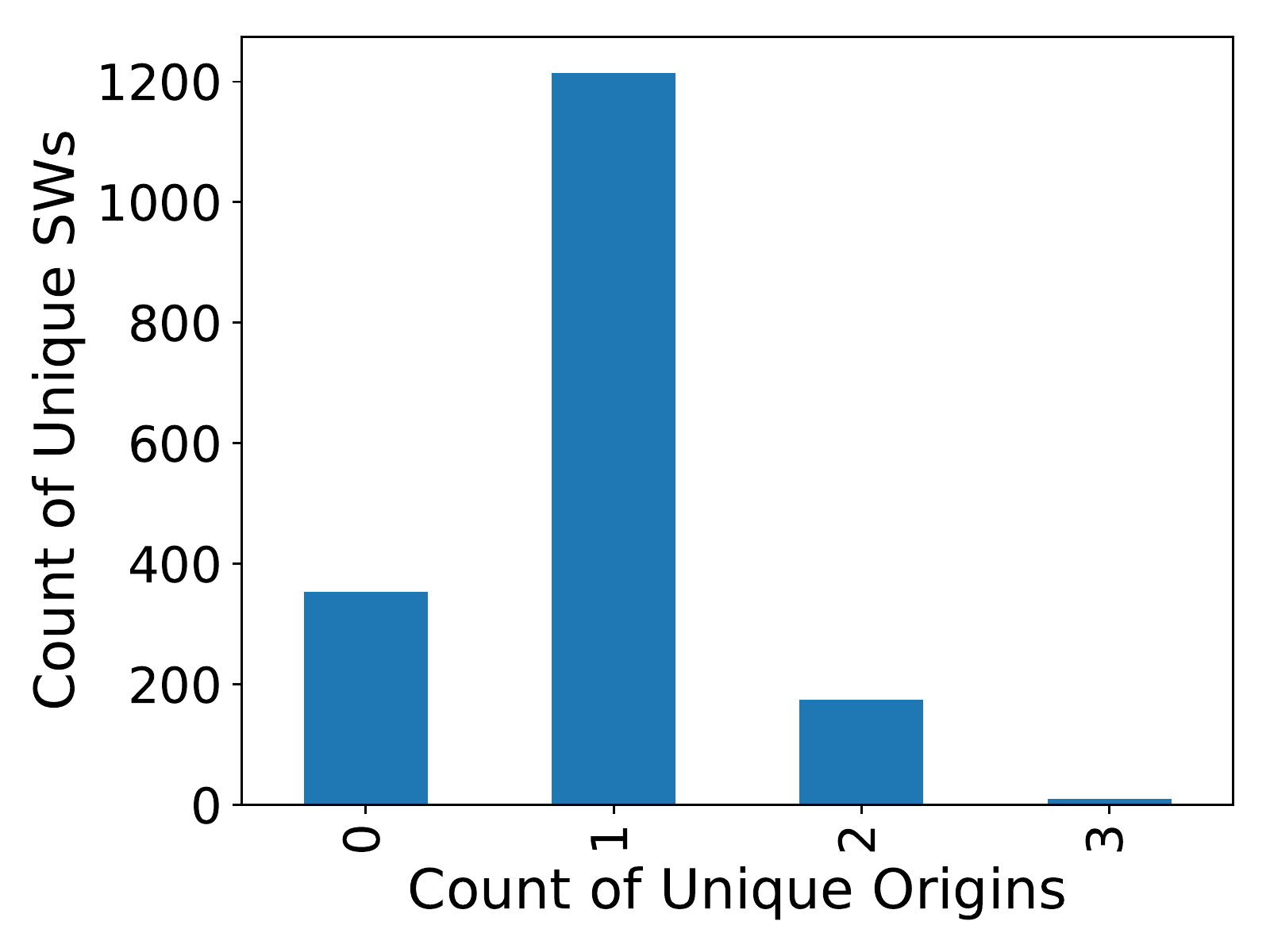}
        \caption{}
        \label{fig:cdf_tp_imports}
    \end{subfigure}%
    \begin{subfigure}{0.48\columnwidth}
    \centering
     \includegraphics[scale=0.4]{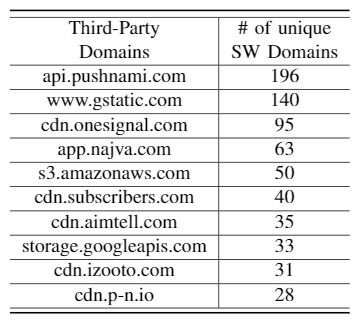}
        \caption{}
        \label{tab:third_party_imports}
    \end{subfigure}
    \caption{Third-Party(TP) Imports (a) Count of TP imported domains per SWs (b) Top 10 TP domains imported by SWs}
\end{figure}

To measure whether it is possible to limit the potential for {\em hijacking}
attacks (see Section~\ref{sec:op_csp}), we analyze the number of third-party
scripts imported by SWs. To this end,
Figure~\ref{fig:cdf_tp_imports} shows the count of third-party scripts imported
per SW, whereas Table~\ref{tab:third_party_imports} shows the top 10 origins
related to imported scripts. 

The vast majority of origins recorded in our logs belong to third-party push services (just among the top 10 origins, 7 are related to web push services). Also, as we can see from Figure~\ref{fig:cdf_tp_imports}, the vast majority of SWs that import third-party code load it from at most one or two origins. Therefore, we believe that the {\em fail-safe defaults} approach we proposed in Section~\ref{sec:op_csp}, whereby the browser should set a {\tt script-src: `self'} default CSP for every service worker, could be implemented with no significant impact on existing SWs. This is because the developers of existing SWs would only need to add one or two authorized origins from which additional SW code can be imported (e.g., in the Apache web server this could be done with a minimal {\tt .htaccess} file associated with the SW file hosted on the website's first-party origin~\cite{htaccess}).

We also measured whether CSPs are currently used in relation to SW code. To this end, we analyzed the HTTP response for every SW file fetched from the 100K Alexa websites. To identify the request (and related response) for a SW file, we look for the {\em Service-Worker} request header, which is explicitly found when a page fetches a SW file. We found that 4.8\% of all SW files have a CSP headers in the response. However, only 0.8\% include the {\tt script-src} directive. Applying {\tt script-src: `self'} as a default policy would reduce the risk of potential hijacking attacks, such as the {\bf XSS} attacks presented in~\cite{Chinprutthiwong2020}.

\subsubsection{Third-Party Code Behavior}

Although the default CSP restrictions discussed above may help to mitigate some
hijacking attacks, such as {\bf XSS} attacks~\cite{Chinprutthiwong2020}, we need
to be mindful that third-party libraries explicitly allowed by web developers
may still behave maliciously. For example, as we studied the results of our
measurements on SW code imports, we found a few cases of potential unauthorized
tracking implemented in third-party libraries. For instance, we found that code
imported from {\em coinPush} 
(a third-party push service) listens to {\em fetch} events, and can
track all URLs visited on the importing website, even though this may not be
necessary to enable the advertised push notification services.
Listing~\ref{code:mal_code} shows the (simplified) code that appears to track
all visited URLs and send them to a remote site. 

Such examples demonstrate that it is possible for third-party services to
potentially abuse their privileged access to SW code, and should therefore be
subjected to stricter default policies by browsers and much more attention by
developers. Therefore, our recommendation is that the browser should implement
more fine-grained SW policies (similar to {\em Feature Policy} for {\tt
iframe}s) and web developers should carefully isolate SWs that import
third-party code by correctly setting their {\em scope}, as we proposed in
Section~\ref{sec:restricting_tpc}.

\begin{lstlisting}[caption={SW Code of 3rd Party Push service },captionpos=b,language=JavaScript, label=code:mal_code]
//Unauthorized Fetch handler to track request URLs
self.addEventListener('fetch', function (e) {
  if (e.request.url.indexOf(location.origin) === 0 && isDocument(e.request)) {
    trackingUrl(e.request.url); }  });
//simplified version of trackingURl method
function trackingUrl(url){
    var body = {registration_id: obj.token,
                sender_id: obj.senderId,
                logs: {url: url, timestamp: timestamp}}
    return fetch(TRACKING_SERVER + '/tracking_url', { method: 'POST', mode: 'no-cors', body: body } 
}
\end{lstlisting}

%% file: tables/alexa_bands.tex
\begin{table}[htbp]
\setlength{\belowcaptionskip}{-5pt} 
\setlength{\abovecaptionskip}{5pt plus 3pt minus 5pt} 
\caption{Ranking bands for Alexa's top 100K websites}
\centering
\label{tab:alexa_Ranking}
\resizebox{\columnwidth}{!}{%

\begin{tabular}{c|c|c|c|c|c|c|c}
\toprule
\hline
\textbf{Ranking Bands} & {Band\_1} & {Band\_2} & {Band\_3} & {Band\_4} & {Band\_5} &{Band\_6} & \multirow{2}{*}{Total}

\\
\cline{1-7}
\textbf{Ranking Range}  & {0-1K} & {1K-5K} & {5K-10K} & {10K-50K} & {50K-100K} & {100K -1M} &
 
\\
\hline
\textbf{\#Registered SWs}  & {160} & {426} & {437} & {1917} & {2369} & {609} & {5918}
 
\\
\hline

\textbf{\#Analyzed SWs}  & {56} & {151} & {145} & {585} & {204} & {609} & {1750}
 
\\
\hline
\bottomrule
\end{tabular}

}
\end{table}

%% file: tables/percentile.tex
\begin{table*}[htbp]
\setlength{\abovecaptionskip}{5pt plus 3pt minus 2pt} 
\caption{Event counts at specific distribution percentiles -- The {\em threshold value} is the percentile value from the corresponding CDF (see Figure~\ref{sec:open_problems}), whereas B-$n$ represents the Alexa ranking band $n$.}
\centering
\label{tab:percentile_bands}
\resizebox{\textwidth}{!}{%
\begin{tabular}{l|l|l|l|l|l|l|l|l|l|l|l|l|l|l|l|l|l|l|l|l|l|l}
\toprule
\hline
\multicolumn{1}{c|}{\multirow{3}{*}{\begin{tabular}[c]{@{}c@{}}Event \\ Count\end{tabular}}} & \multirow{3}{*}{\begin{tabular}[c]{@{}c@{}} \\ No. of \\ SW \\ Origins\end{tabular}} & \multicolumn{21}{c}{Number of SW Origins above threahold value} \\ \cline{3-23} 
\multicolumn{1}{c|}{} &  & \multicolumn{7}{c|}{\begin{tabular}[c]{@{}c@{}} 90\%\end{tabular}} & \multicolumn{7}{c|}{\begin{tabular}[c]{@{}c@{}}95\%\end{tabular}} & \multicolumn{7}{c}{\begin{tabular}[c]{@{}c@{}}99\%\end{tabular}} \\ \cline{3-23} 
\multicolumn{1}{c|}{} &  & \multicolumn{1}{c|}{\begin{tabular}[c]{@{}c@{}}Threshold\\ Value\end{tabular}} & {B-1} & {B-2} & {B-3} & {B-4} & {B-5} & {B-6} & \multicolumn{1}{c|}{\begin{tabular}[c]{@{}c@{}}Threshold\\ Value\end{tabular}} & {B-1} & {B-2} & {B-3} & {B-4} & {B-5} & {B-6} & \multicolumn{1}{c|}{\begin{tabular}[c]{@{}c@{}}Threshold\\ Value\end{tabular}} & {B-1} & {B-2} & {B-3} & {B-4} & {B-5} & B-6 \\ \hline
Push Count per Hour & 518 & {14} & {0} & {1} & {5} & {11} & {10} & {22} & {22} & {0} & {1} & {3} & {7} & {5} & {8} & {43} & {0} & {0} & {2} & {6} & {2} & 5 \\ \hline
Push Count per Day & 518 & {38} & {1} & {2} & {5} & {13} & {9} & {43} & {88} & {1} & {1} & {3} & {8} & {4} & {11} & {392} & {0} & {0} & {0} & {3} & {0} & 1 \\ \hline
Third Party Fetch Count per Activation  & 416 & {1} & {12} & {10} & {13} & {35} & {30} & {44} & {2} & {8} & {8} & {10} & {25} & {26} & {36} & {6} & {5} & {4} & {8} & {12} & {19} & 17 \\ \hline
SW Execution Time Per Activation & 761 & {3} & {2} & {1} & {6} & {16} & {16} & {38} & {3} & {2} & {1} & {6} & {16} & {16} & {38} & {5} & {1} & {1} & {2} & {5} & {5} & 6 \\ \hline
SW Execution Time Per Day & 761 & {64} & {1} & {2} & {7} & {9} & {10} & {33} & {90} & {1} & {0} & {3} & {6} & {4} & {26} & {146} & {1} & {0} & {2} & {3} & {2} & 9 \\ \hline

\bottomrule
\end{tabular}

}
\end{table*}

%% file: policies_implementation.tex
\section{Implementing New SW Policies}
\label{sec:policy_impl}

To demonstrate that implementing the new policies proposed in
Section~\ref{sec:open_problems} is possible with reasonable engineering effort,
in the following we discuss our own proof-of-concept implementation in the
Chromium (v84.0.4147.121) browser of some of those policies. We plan to release our source code after publication.

\begin{lstlisting}[caption={Template for count-based policies},captionpos=b,language=C, label=code:sw_policy_template]
{   "name": <policy-name>,    
    "severity": <value>,  
    "threshold": <threshold>, 
    "duration_in_minutes":<duration>  }
\end{lstlisting}  

To implement the new policies, we developed a new class called \texttt{SWPolicies}
within the {\em Blink} rendering engine. In case of policies concerned with
limiting the frequency of events that activate a SW, we follow a template
similar to the one shown in Listing~\ref{code:sw_policy_template}. Each time an
event such as {\em push} or {\em fetch} occurs, we invoke a corresponding method
to update the related counter (e.g., \textit{push\_count\_per\_hour}) and check
if the count falls within a predefined \textit{threshold}, which could be
selected based on the trade-offs we discussed in our SW behavior measurements
results (see Section~\ref{sec:measurements}). When a policy violation occurs
(i.e., the \textit{threshold} is exceeded), we log the violation and increase a
\textit{severity} indicator. Then, if the severity level reaches a predefined
maximum value, the browser immediately terminates the SW (and could also
deregister it, if the user engagement score for the SW's origin is very low).

\begin{lstlisting}[caption={Example code for limiting SW execution time},captionpos=b,language=C, label=code:sw_policy_time]
void SWPolicies::OnSWActivated(ExecutionContext* ex)
{ //start timer
  swpolicy_info->sw_timer->Start(
      FROM_HERE, kServiceWorkerRunningDelay, 
      base::BindOnce(&SWPolicies::OnSWTimeout,
      base::Unretained(this), To<SWGlobalScope>(ex)));
}
void SWPolicies::OnSWTimeout(SWGlobalScope* gs)
{ // immediately terminate SW
  gs->SetIdleDelay(base::TimeDelta::FromSeconds(0));
}
\end{lstlisting}   

As a simplified example of how to terminate a SW, to stop a SW that exceeds a
given execution time we can start a timer whenever the service worker is
activated, as shown in Listing~\ref{code:sw_policy_time}, and attach a callback
method that will be called once the timer expires. At this point, the callback
can check the state of the SW and terminate it (if it is still running) by
calling Blink's \texttt{SetIdleDelay} method with delay set to 0 seconds.

We use an approach similar to that described above to implement and enforce the policies proposed in Section~\ref{sec:op_limiting_exe}. Furthermore, we tested these policies against a number of SW attacks (using the approaches we described in Section~\ref{sec:attacks}) that attempt to perform DDoS attacks, cryptomining, notification spam, etc., and verified that we are indeed able to greatly throttle such attacks, rendering them ineffective.

\subsection{Discussion}
\label{sec:discussion}
As we consider the implementation of new browser policies that would restrict SW
behaviors, as discussed earlier, we should carefully consider how they could
impact legitimate SWs. In our measurements (Section~\ref{sec:measurements}), we
showed that it would be possible to find enforcement thresholds whose effect is
to greatly limit abuse while interfering with the behavior of only few actual
legitimate SWs. In addition, we should consider that the implementation of the
proposed policies could include a customizable allow-list that can be pre-populated by the
browser vendor and extended with help from the user, if preferred. For instance, consider
the limits on the frequency of push notifications proposed in
Section~\ref{sec:op_limiting_exe}. If a popular website (e.g., a social media
platform) legitimately needs to send a large number of notifications (e.g.,
many tens or hundreds of notifications per day), such an application could be
added to this allow-list.

The effect of the proposed policies against continuous execution attacks
(Section~\ref{sec:open_problems}) is that they may also limit the execution time
for a small fraction of legitimate SWs. However, we should notice that browsers
already implement a mechanism to terminate a SW's execution after a certain
amount of execution time. Therefore, SW developers already need to take into
account that their code could be forced into an idle state. Unfortunately, as we
demonstrated in Section~\ref{sec:attacks}, attackers can trick the browser into
executing SWs for much longer than it would be otherwise allowed, which
motivates the new policies we proposed Section~\ref{sec:open_problems}.
Ultimately, we believe that limiting the execution time of SWs would have a
small to negligible impact on legitimate SW code, while drastically reducing the
risk for SW abuse. Additionally, before enforcing the new policies, browsers
could grant a grace period during which an alert is issued every time a SW
policy is violated without strictly enforcing the policy itself. During this
transition period, developers will then have the time to adjust their SW code to
make sure the new policies are not violated moving forward.

In general, because the vast majority of legitimate web applications do not require a completely
unfettered access to push events, web notifications, background third-party
requests, etc., as shown in Section~\ref{sec:measurements}, and considering the
potential damage that SW abuse could cause given its powerful features (see
Section~\ref{sec:attacks}), we believe browsers should follow an approach akin
to the {\em least privileges} principle as much as possible and limit those
and other SW privileges.

%% file: relatedwork.tex
\section{Related Work}
\label{sec:relatedwork}
Throughout the paper we have discussed a number of previous works that focus
primarily on attacks against SWs or in which SWs play a fundamental role. In
this section, we briefly discuss other works related to multiple aspects of SW
security, including some additional attacks and mitigation measures.  

New, recently published studies~\cite{cache_woot, Watanabe2020MeltingPO} discuss other attacks that leverage SW features. In~\cite{cache_woot}, Squarcina
et al. describe a powerful DOM-based XSS attack that could be used to inject code
into a script cached by a SW, through which the injected code remains persistent
in the client's machine for a longer duration compared to traditional XSS
attacks. This attack is similar to cache poisoning attacks for the {\em
Browser Cache}~\cite{cache_1,persistent_cache}, {\em Client
Storage}~\cite{client_storage_poisoning} and {\em Web Cache}~\cite{cache_poison,
cache_usenix} mechanisms. Unlike attacks discussed in Section~\ref{sec:attacks}, the role of SWs in this attack is limited to serving already poisoned cache resources. However, it still highlights the need for stricter access restrictions similar to our proposed mitigation of isolating SW cache access (Section~\ref{sec:mitigations}).


In~\cite{Watanabe2020MeltingPO}, the authors discuss a number of security issues
related to rehosted websites and describe a persistent man-in-the-middle attack
implemented by leveraging SWs. Since all rehosted websites may be placed under
the same origin as the attacker site that registers a SW, the SW could gain
control of all resources under the rehosted origin, rendering the
Same-Origin-Policy for SWs ineffective. Mitigating this attack likely requires
defense measures at the rehosting level, as the attack violates a number of
security policies.

Soon after SWs were introduced by browsers, a few studies~\cite{sw_timing_attack,cache_leakage} discussed the possibility
of side-channel attacks that leverage SW features.
More works~\cite{ext_security, pantelaios2020you, botnet_extension,thomas2015ad,
lauinger2018thou} are dedicated to measuring privacy leakage due to third-party
code, such as extensions, external libraries and ad injections.
In~\cite{meiser2021careful}, the authors discuss issues related to blind trust
between cross origins and explain the extent of damage it could lead to. Other
works that focus on client-side web security are~\cite{client_web,
jueckstock2019visiblev8,schwarz2018javascript}, which include policy-based
access to restict Javascipt APIs(such as Performance API) to mitigate timing
based side-channel attacks. Similarly, \cite{policy_security} presents a
cost-benefit analysis to enable restrictions per site without affecting its
legitimate use. 
In \cite{fine_grained}, Jackson and Barth argue that the concept of
``fine-grained origins'' (FGO) is a flawed solution to curb origin
contamination. In this paper (Section~\ref{sec:restricting_tpc}), we discussed
the use of scopes as a way to isolate third-party SWs and proposed additional
measures to restrict the capabilities of third-party SWs to mitigate possible
origin contamination issues for SW scripts.


Other systematization of knowledge papers that consider both attacks and
mitigation measures have focused on web security~\cite{sok_browser,sok_all,
sok_mitm}, mobile security~\cite{sok_android}, and IoT security~\cite{sok_iot}.
Our work is different because we focus on attacks on Service Workers
specifically, discuss existing mitigations, present a timeline of when attacks
and mitigations were introduced, present open security problems, and proposed
new policies that browsers could adopt to limit the damage that SW abuse could
cause. 

%% file: conclusion.tex
\section{Conclusion}
\label{sec:conclusion}
 
In this paper, we reproduced and analyzed known attack vectors related to
Service Workers and explored new abuse paths that have not previously been
considered. We systematized the attacks into different categories, and analyzed
whether, how, and when these attacks have been published and mitigated by
different browser vendors. Then, we discussed a number of open SW security
problems that are currently unmitigated, and propose SW behavior monitoring
approaches and new browser policies that we believe should be implemented by
browsers to further improve SW security. Furthermore, we implement a
proof-of-concept version of several policies in the Chromium code base, and also
measure the behavior of SWs used by highly popular web applications with respect
to these new policies. Our measurements show that it is feasible to implement
and enforce stricter SW security policy without a significant impact on most
legitimate production SWs.

%% file: appendix.tex
\section{Appendix}
\label{sec:appendix}

\subsection{Additional Example Code Snippets}
\begin{minipage}{\linewidth}

\begin{lstlisting}[caption={Example code to avoid showing notifications on push events},language=JavaScript, label=code:show_not_code]

    self.addEventListener('push', async function (event) {
        // listen to push event and perform any computation here
        // increase push count
        push_count +=1
        // do not call ShowNotification() API
        
        if(push_count>10){
        //renew subscription
            var options = {
              userVisibleOnly: true,
              applicationServerKey: <applicationServerKey>
            };
            self.registration.pushManager.getSubscription().then(function(subscription) {
                subscription.unsubscribe().then(function(successful) {
                  // You've successfully unsubscribed
                  // subscribe again
                  self.registration.pushManager.subscribe(options)
                }).catch(function(e) {
                  // Unsubscribe failed
            })
        }
    });
\end{lstlisting}

\begin{lstlisting}[caption={Hijacking SW code from an extension},captionpos=b,language=JavaScript, label=code:ext_code]
    function modifySW(url, reqId) {
          let filter = browser.webRequest.filterResponseData(reqId);
          ...
          filter.ondata = event => {
              let str = decoder.decode(event.data, {stream: true});
              let code_snippet = "self.addEventListener('push', async function (event) {
               console.log('Extension:: Received push')} );"
              filter.write(encoder.encode(malicious_code+'\n'+str));
              filter.disconnect();
          }
    }
\end{lstlisting}

\end{minipage}
\begin{lstlisting}[caption={Reusing the same notification for multiple push messages},captionpos=b,language=JavaScript, label=code:notification_reuse]
    self.addEventListener('push', async function (event) {
        var notificationTitle = 'Same Notification!';
        var notificationOptions = {
            body: event.data.text(),
            //same tag used for all incoming push events 
            tag: 'notification-update-tag'
        };
        // replaces current displayed notification 
        // with new notification due to using the same tag
        self.registration.showNotification(notificationTitle, notificationOptions)
    }
\end{lstlisting}

\begin{figure*}[ht]
    \begin{center}
        \begin{subfigure}{0.9\columnwidth}
        \begin{center}
            \includegraphics[scale=0.5]{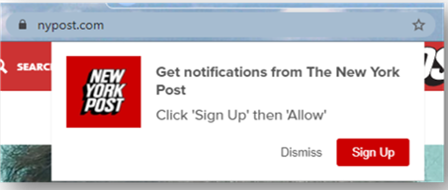}
            \caption{JavaScript-based prompt}
        \end{center}
        \end{subfigure} \\[3ex]
        \begin{subfigure}{0.9\columnwidth}
        \begin{center}
            \includegraphics[scale=0.5]{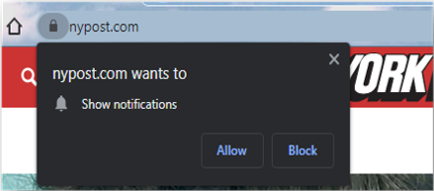}
            \caption{Browser native prompt, after user clicked on `Sign Up'}
        \end{center}
        \end{subfigure} 
        \caption{Example of double permission prompt in use on a popular website.}
        \label{fig:double_perm}
    \end{center}
\end{figure*}

%% file: main.bbl
\begin{thebibliography}{10}
\providecommand{\url}[1]{#1}
\csname url@samestyle\endcsname
\providecommand{\newblock}{\relax}
\providecommand{\bibinfo}[2]{#2}
\providecommand{\BIBentrySTDinterwordspacing}{\spaceskip=0pt\relax}
\providecommand{\BIBentryALTinterwordstretchfactor}{4}
\providecommand{\BIBentryALTinterwordspacing}{\spaceskip=\fontdimen2\font plus
\BIBentryALTinterwordstretchfactor\fontdimen3\font minus
  \fontdimen4\font\relax}
\providecommand{\BIBforeignlanguage}[2]{{%
\expandafter\ifx\csname l@#1\endcsname\relax
\typeout{** WARNING: IEEEtran.bst: No hyphenation pattern has been}%
\typeout{** loaded for the language `#1'. Using the pattern for}%
\typeout{** the default language instead.}%
\else
\language=\csname l@#1\endcsname
\fi
#2}}
\providecommand{\BIBdecl}{\relax}
\BIBdecl

\bibitem{sw_intro}
``Mdn web docs: Service worker,''
  \url{https://developers.google.com/web/fundamentals/primers/service-workers}.

\bibitem{sw_power}
``The chromium projects,''
  \url{https://www.chromium.org/Home/chromium-security/prefer-secure-origins-for-powerful-new-features}.

\bibitem{pwa_intro}
``Web.dev: Progressive web apps,'' \url{https://web.dev/progressive-web-apps/}.

\bibitem{push_api}
``Mdn web docs:push api,''
  \url{https://developer.mozilla.org/en-US/docs/Web/API/Push_API}.

\bibitem{masterofweb}
P.~Papadopoulos, P.~Ilia, M.~Polychronakis, E.~Markatos, S.~Ioannidis, and
  G.~Vasiliadis, ``Master of web puppets: Abusing web browsers for persistent
  and stealthy computation,'' \emph{ArXiv}, vol. abs/1810.00464, 2019.

\bibitem{pridepwa}
J.~Lee, H.~Kim, J.~Park, I.~Shin, and S.~Son, ``Pride and prejudice in
  progressive web apps: Abusing native app-like features in web applications,''
  in \emph{Proceedings of the 2018 ACM SIGSAC Conference on Computer and
  Communications Security}, 2018, pp. 1731--1746.

\bibitem{Chinprutthiwong2020}
\BIBentryALTinterwordspacing
P.~Chinprutthiwong, R.~Vardhan, G.~Yang, and G.~Gu, ``Security study of service
  worker cross-site scripting.'' in \emph{Annual Computer Security Applications
  Conference}, ser. ACSAC '20.\hskip 1em plus 0.5em minus 0.4em\relax New York,
  NY, USA: Association for Computing Machinery, 2020, p. 643–654. [Online].
  Available: \url{https://doi.org/10.1145/3427228.3427290}
\BIBentrySTDinterwordspacing

\bibitem{karami2021awakening}
S.~Karami, P.~Ilia, and J.~Polakis, ``Awakening the web’s sleeper agents:
  Misusing service workers for privacy leakage,'' in \emph{Network and
  Distributed System Security Symposium (NDSS)}, 2021.

\bibitem{subramani2020}
\BIBentryALTinterwordspacing
K.~Subramani, X.~Yuan, O.~Setayeshfar, P.~Vadrevu, K.~H. Lee, and R.~Perdisci,
  ``When push comes to ads: Measuring the rise of (malicious) push
  advertising,'' in \emph{Proceedings of the ACM Internet Measurement
  Conference}, ser. IMC '20.\hskip 1em plus 0.5em minus 0.4em\relax New York,
  NY, USA: Association for Computing Machinery, 2020, p. 724–737. [Online].
  Available: \url{https://doi.org/10.1145/3419394.3423631}
\BIBentrySTDinterwordspacing

\bibitem{attacks_repo}
``Sw abuse demos,'' \url{https://demopwa.github.io/sw_index}.

\bibitem{sw_reg}
``Mdn web docs: Serviceworkerregistration,''
  \url{https://developer.mozilla.org/en-US/docs/Web/API/ServiceWorkerRegistration}.

\bibitem{push_manager_api}
``Mdn web docs:pushmanager api,''
  \url{https://developer.mozilla.org/en-US/docs/Web/API/PushManager/subscribe}.

\bibitem{push_subscription_api}
``Mdn web docs:push subscription api,''
  \url{https://developer.mozilla.org/en-US/docs/Web/API/PushSubscription/getKey}.

\bibitem{periodic_sync_api}
``Mdn web docs : Periodic background synchronization api,''
  \url{https://developer.mozilla.org/en-US/docs/Web/API/Web_Periodic_Background_Synchronization_API}.

\bibitem{sync_concerns}
``Google groups: Intent to ship: Periodic background sync,''
  \url{https://groups.google.com/a/chromium.org/g/blink-dev/c/KSJViFp3hMc/m/6gVYzjg_BAAJ?pli=1}.

\bibitem{periodic_sync_rules}
``Web.dev : Richer offline experiences with periodic background sync api,''
  \url{https://web.dev/periodic-background-sync/}.

\bibitem{periodic_sync_security}
``W3c periodic background sync specification,''
  \url{https://wicg.github.io/background-sync/spec/PeriodicBackgroundSync-index.html\#security
  }.

\bibitem{sync_issue}
``Periodic background sync has serious security risks,'' \url{
  https://github.com/WICG/background-sync/issues/169}.

\bibitem{sw_policies}
``Service worker security policies,''
  \url{https://chromium.googlesource.com/chromium/src/+/master/docs/security/service-worker-security-faq.md}.

\bibitem{sw_security}
``W3c: Service worker,''
  \url{https://www.w3.org/TR/service-workers/#security-considerations}.

\bibitem{sync_manager_api}
``Mdn web docs: Syncmanager api,''
  \url{https://developer.mozilla.org/en-US/docs/Web/API/SyncManager}.

\bibitem{pantelaios2020you}
N.~Pantelaios, N.~Nikiforakis, and A.~Kapravelos, ``You've changed: Detecting
  malicious browser extensions through their update deltas,'' in
  \emph{Proceedings of the 2020 ACM SIGSAC Conference on Computer and
  Communications Security}, 2020, pp. 477--491.

\bibitem{update_issue_firefox}
``Bugzilla : Self-update service worker to stay alive,''
  \url{https://bugzilla.mozilla.org/show_bug.cgi?id=1432846}.

\bibitem{google_usenix}
\BIBentryALTinterwordspacing
I.~Bilogrevic, B.~Engedy, J.~Porter, N.~Taft, K.~Hasanbega, A.~Paseltiner,
  H.~Lee, E.~Jung, M.~Watkins, P.~McLachlan, and J.~James, ``"shhh...be quiet!"
  reducing the unwanted interruptions of notification permission prompts on
  chrome,'' in \emph{30th {USENIX} Security Symposium ({USENIX} Security 21)},
  Vancouver, B.C., 2021. [Online]. Available:
  \url{https://www.usenix.org/conference/usenixsecurity21/presentation/bilogrevic}
\BIBentrySTDinterwordspacing

\bibitem{quiet_ui}
``Quieter ui for notificationss,''
  \url{https://blog.chromium.org/2020/01/introducing-quieter-permission-ui-for.html}.

\bibitem{quiet_ui_firefox}
``Mozilla : Web push notifications,''
  \url{https://support.mozilla.org/en-US/kb/push-notifications-firefox\#w_how-does-it-work}.

\bibitem{Bouwman2020}
\BIBentryALTinterwordspacing
X.~Bouwman, H.~Griffioen, J.~Egbers, C.~Doerr, B.~Klievink, and M.~van Eeten,
  ``A different cup of {TI}? the added value of commercial threat
  intelligence,'' in \emph{29th {USENIX} Security Symposium ({USENIX} Security
  20)}.\hskip 1em plus 0.5em minus 0.4em\relax {USENIX} Association, Aug. 2020,
  pp. 433--450. [Online]. Available:
  \url{https://www.usenix.org/conference/usenixsecurity20/presentation/bouwman}
\BIBentrySTDinterwordspacing

\bibitem{double_permission_chrome}
``Webfundamentals: Permission ux,''
  \url{https://developers.google.com/web/fundamentals/push-notifications/permission-ux\#double_permission}.

\bibitem{double_permission_onesignal}
``Investigating the chrome ux report: How to boost web push opt-in rates,''
  \url{https://onesignal.com/blog/boost-your-web-push-opt-in-rates/}.

\bibitem{phani_imc}
\BIBentryALTinterwordspacing
P.~Vadrevu and R.~Perdisci, ``What you see is not what you get: Discovering and
  tracking social engineering attack campaigns,'' in \emph{Proceedings of the
  Internet Measurement Conference}, ser. IMC '19.\hskip 1em plus 0.5em minus
  0.4em\relax New York, NY, USA: Association for Computing Machinery, 2019, p.
  308–321. [Online]. Available: \url{https://doi.org/10.1145/3355369.3355600}
\BIBentrySTDinterwordspacing

\bibitem{site_engagement}
``Chromium projects : Site engagement,''
  \url{https://sites.google.com/a/chromium.org/dev/developers/design-documents/site-engagement}.

\bibitem{Nikiforakis2012}
\BIBentryALTinterwordspacing
N.~Nikiforakis, L.~Invernizzi, A.~Kapravelos, S.~Van~Acker, W.~Joosen,
  C.~Kruegel, F.~Piessens, and G.~Vigna, ``You are what you include:
  Large-scale evaluation of remote javascript inclusions,'' in
  \emph{Proceedings of the 2012 ACM Conference on Computer and Communications
  Security}, ser. CCS '12.\hskip 1em plus 0.5em minus 0.4em\relax New York, NY,
  USA: Association for Computing Machinery, 2012, p. 736–747. [Online].
  Available: \url{https://doi.org/10.1145/2382196.2382274}
\BIBentrySTDinterwordspacing

\bibitem{Weissbacher2014}
M.~Weissbacher, T.~Lauinger, and W.~Robertson, ``Why is csp failing? trends and
  challenges in csp adoption,'' in \emph{Research in Attacks, Intrusions and
  Defenses}, A.~Stavrou, H.~Bos, and G.~Portokalidis, Eds.\hskip 1em plus 0.5em
  minus 0.4em\relax Cham: Springer International Publishing, 2014, pp.
  212--233.

\bibitem{eval_worker}
``Mdn web docs: Using web workers,''
  \url{https://developer.mozilla.org/en-US/docs/Web/API/Web_Workers_API/Using_web_workers\#content_security_policy}.

\bibitem{worker_src}
``Mdn web docs: Csp worker-src,''
  \url{https://developer.mozilla.org/en-US/docs/Web/HTTP/Headers/Content-Security-Policy/worker-src}.

\bibitem{cookie_store}
``W3c cookie store api,'' \url{https://wicg.github.io/cookie-store/}.

\bibitem{cache_api}
``Mdn web docs: Cache api,''
  \url{https://developer.mozilla.org/en-US/docs/Web/API/Cache}.

\bibitem{appcache_api}
``Appcache api,''
  \url{https://blog.chromium.org/2020/01/appcache-scope-restricted.html}.

\bibitem{feature_policy}
``Mdn web docs: Iframe,''
  \url{https://developer.mozilla.org/en-US/docs/Web/HTTP/Feature_Policy/Using_Feature_Policy}.

\bibitem{alexa}
``Alexa top sites,'' \url{https://www.alexa.com/topsites}.

\bibitem{puppeteer}
Google, ``Puppeteer: Chormium browser automation tool,''
  \url{http://liwc.wpengine.com/compare-dictionaries/}, 2019, {(Last accessed
  Nov.11, 2019)}.

\bibitem{htaccess}
``Adding csp using .htaccess,''
  \url{https://content-security-policy.com/examples/htaccess/}.

\bibitem{cache_woot}
M.~Squarcina, D.~F. Some, S.~Calzavara, and M.~Maffei, ``The remote on the
  local: Exacerbating web attacks via service workers caches,'' ser. USENIX
  WOOT '21.

\bibitem{Watanabe2020MeltingPO}
T.~Watanabe, E.~Shioji, M.~Akiyama, and T.~Mori, ``Melting pot of origins:
  Compromising the intermediary web services that rehost websites,'' in
  \emph{NDSS}, 2020.

\bibitem{cache_1}
Y.~Jia, Y.~Chen, X.~Dong, P.~Saxena, J.~Mao, and Z.~Liang,
  ``Man-in-the-browser-cache: persisting https attacks via browser cache
  poisoning,'' \emph{computers \& security}, vol.~55, pp. 62--80, 2015.

\bibitem{persistent_cache}
M.~Vallentin and Y.~Ben-David, ``Persistent browser cache poisoning,'' 2010.

\bibitem{client_storage_poisoning}
M.~Steffens, C.~Rossow, M.~Johns, and B.~Stock, ``Don’t trust the locals:
  Investigating the prevalence of persistent client-side cross-site scripting
  in the wild.'' 2019.

\bibitem{cache_poison}
\BIBentryALTinterwordspacing
J.~Kettle, ``Practical web cache poisoning,'' ser. BlackHat 2018. [Online].
  Available:
  \url{https://portswigger.net/research/practical-web-cache-poisoning}
\BIBentrySTDinterwordspacing

\bibitem{cache_usenix}
S.~A. Mirheidari, S.~Arshad, K.~Onarlioglu, B.~Crispo, E.~Kirda, and
  W.~Robertson, ``Cached and confused: Web cache deception in the wild,'' in
  \emph{29th $\{$USENIX$\}$ Security Symposium ($\{$USENIX$\}$ Security 20)},
  2020, pp. 665--682.

\bibitem{sw_timing_attack}
\BIBentryALTinterwordspacing
T.~Van~Goethem, W.~Joosen, and N.~Nikiforakis, ``The clock is still ticking:
  Timing attacks in the modern web,'' in \emph{Proceedings of the 22nd ACM
  SIGSAC Conference on Computer and Communications Security}, ser. CCS
  '15.\hskip 1em plus 0.5em minus 0.4em\relax New York, NY, USA: Association
  for Computing Machinery, 2015, p. 1382–1393. [Online]. Available:
  \url{https://doi.org/10.1145/2810103.2813632}
\BIBentrySTDinterwordspacing

\bibitem{cache_leakage}
T.~Van~Goethem, M.~Vanhoef, F.~Piessens, and W.~Joosen, ``Request and conquer:
  Exposing cross-origin resource size,'' in \emph{25th $\{$USENIX$\}$ Security
  Symposium ($\{$USENIX$\}$ Security 16)}, 2016, pp. 447--462.

\bibitem{ext_security}
\BIBentryALTinterwordspacing
O.~Starov and N.~Nikiforakis, ``Extended tracking powers: Measuring the privacy
  diffusion enabled by browser extensions,'' in \emph{Proceedings of the 26th
  International Conference on World Wide Web}, ser. WWW '17.\hskip 1em plus
  0.5em minus 0.4em\relax Republic and Canton of Geneva, CHE: International
  World Wide Web Conferences Steering Committee, 2017, p. 1481–1490.
  [Online]. Available: \url{https://doi.org/10.1145/3038912.3052596}
\BIBentrySTDinterwordspacing

\bibitem{botnet_extension}
R.~{Perrotta} and F.~{Hao}, ``Botnet in the browser: Understanding threats
  caused by malicious browser extensions,'' \emph{IEEE Security Privacy},
  vol.~16, no.~4, pp. 66--81, 2018.

\bibitem{thomas2015ad}
K.~Thomas, E.~Bursztein, C.~Grier, G.~Ho, N.~Jagpal, A.~Kapravelos, D.~McCoy,
  A.~Nappa, V.~Paxson, P.~Pearce \emph{et~al.}, ``Ad injection at scale:
  Assessing deceptive advertisement modifications,'' in \emph{2015 IEEE
  Symposium on Security and Privacy}.\hskip 1em plus 0.5em minus 0.4em\relax
  IEEE, 2015, pp. 151--167.

\bibitem{lauinger2018thou}
T.~Lauinger, A.~Chaabane, S.~Arshad, W.~Robertson, C.~Wilson, and E.~Kirda,
  ``Thou shalt not depend on me: Analysing the use of outdated javascript
  libraries on the web,'' \emph{arXiv preprint arXiv:1811.00918}, 2018.

\bibitem{meiser2021careful}
G.~Meiser, P.~Laperdrix, and B.~Stock, ``Careful who you trust: Studying the
  pitfalls of cross-origin communication,'' \emph{ACM AsiaCCS}, 2021.

\bibitem{client_web}
\BIBentryALTinterwordspacing
B.~Stock, M.~Johns, M.~Steffens, and M.~Backes, ``How the web tangled itself:
  Uncovering the history of client-side web (in)security,'' in \emph{26th
  {USENIX} Security Symposium ({USENIX} Security 17)}.\hskip 1em plus 0.5em
  minus 0.4em\relax Vancouver, BC: {USENIX} Association, Aug. 2017, pp.
  971--987. [Online]. Available:
  \url{https://www.usenix.org/conference/usenixsecurity17/technical-sessions/presentation/stock}
\BIBentrySTDinterwordspacing

\bibitem{jueckstock2019visiblev8}
J.~Jueckstock and A.~Kapravelos, ``Visiblev8: In-browser monitoring of
  javascript in the wild,'' in \emph{Proceedings of the Internet Measurement
  Conference}, 2019, pp. 393--405.

\bibitem{schwarz2018javascript}
M.~Schwarz, M.~Lipp, and D.~Gruss, ``Javascript zero: Real javascript and zero
  side-channel attacks.'' in \emph{NDSS}, 2018.

\bibitem{policy_security}
\BIBentryALTinterwordspacing
P.~Snyder, C.~Taylor, and C.~Kanich, ``Most websites don't need to vibrate: A
  cost-benefit approach to improving browser security,'' in \emph{Proceedings
  of the 2017 ACM SIGSAC Conference on Computer and Communications Security},
  ser. CCS '17.\hskip 1em plus 0.5em minus 0.4em\relax New York, NY, USA:
  Association for Computing Machinery, 2017, p. 179–194. [Online]. Available:
  \url{https://doi.org/10.1145/3133956.3133966}
\BIBentrySTDinterwordspacing

\bibitem{fine_grained}
C.~Jackson, ``Beware of finer-grained origins.''

\bibitem{sok_browser}
R.~Rogowski, M.~Morton, F.~Li, F.~Monrose, K.~Z. Snow, and M.~Polychronakis,
  ``Revisiting browser security in the modern era: New data-only attacks and
  defenses,'' in \emph{2017 IEEE European Symposium on Security and Privacy
  (EuroS\&P)}.\hskip 1em plus 0.5em minus 0.4em\relax IEEE, 2017, pp. 366--381.

\bibitem{sok_all}
A.~Bulazel and B.~Yener, ``A survey on automated dynamic malware analysis
  evasion and counter-evasion: Pc, mobile, and web,'' in \emph{Proceedings of
  the 1st Reversing and Offensive-oriented Trends Symposium}, 2017, pp. 1--21.

\bibitem{sok_mitm}
M.~Conti, N.~Dragoni, and V.~Lesyk, ``A survey of man in the middle attacks,''
  \emph{IEEE Communications Surveys \& Tutorials}, vol.~18, no.~3, pp.
  2027--2051, 2016.

\bibitem{sok_android}
Y.~Acar, M.~Backes, S.~Bugiel, S.~Fahl, P.~McDaniel, and M.~Smith, ``Sok:
  Lessons learned from android security research for appified software
  platforms,'' in \emph{2016 IEEE Symposium on Security and Privacy
  (SP)}.\hskip 1em plus 0.5em minus 0.4em\relax IEEE, 2016, pp. 433--451.

\bibitem{sok_iot}
O.~Alrawi, C.~Lever, M.~Antonakakis, and F.~Monrose, ``Sok: Security evaluation
  of home-based iot deployments,'' in \emph{2019 IEEE symposium on security and
  privacy (sp)}.\hskip 1em plus 0.5em minus 0.4em\relax IEEE, 2019, pp.
  1362--1380.

\end{thebibliography}
